\documentclass[letterpaper,dvipdfm]{JHEP3}
\title{Phase Transitions and Moduli Space Topology}
\author{D. D. Ferrante\\
  Department of Physics, Syracuse University\\
  Syracuse --- NY. 13244. USA.}
\author{G. S. Guralnik\\
  Department of Physics, Brown University\\
  Providence --- RI. 02912. USA.}
\preprint{SU-4252-878 \\ BROWN-HET-1564}
\abstract{
  By means of an appropriate re-scaling of the metric in a Lagrangian, we are
  able to reduce it to a kinetic term only. This form enables us to examine the
  extended complexified solution set (complex moduli space) of field theories by
  finding all possible geodesics of this metric. This new geometrical standpoint 
  sheds light on some foundational issues of QFT and brings to the forefront
  non-perturbative core aspects of field theory. In this process, we show that
  different phases of the theory are topologically inequivalent, i.e., their
  moduli space has distinct topologies. Moreover, the different phases are
  related by ``duality transformations'', which are established by the modular
  structure of the theory. In conclusion, after the topological structure is
  elucidated, it is possible to use the Euler Characteristic in order to
  topologically quantize the theory, in resonance with the content of the
  Atiyah-Singer Index theorem.
}
\keywords{Spontaneous Symmetry Breaking, Phase Transitions, Nonperturbative
  Effects, Morse Theory, Higgs Bundle, Moduli Space Topology}
\usepackage[T1]{fontenc}
\usepackage{ae,aecompl}
\usepackage{lmodern}
\usepackage{microtype}
\usepackage{ucs}
\usepackage[utf8x]{inputenc}
\usepackage{graphicx}
\usepackage{xspace}
\usepackage{amsmath,amsfonts,amstext,amssymb,amsbsy,amsopn,amsthm,eucal}
\DeclareMathAlphabet{\mathpzc}{T1}{pzc}{m}{it} 
\usepackage{mathrsfs}
\usepackage{wasysym}
\usepackage{empheq}
\usepackage{dsfont}
\usepackage{pifont}

\DeclareMathOperator{\tr}{tr}

\DeclareMathOperator{\Res}{Res}
\DeclareMathOperator{\ed}{d\!}

\DeclareMathOperator{\re}{\mathsf{Re}}

\DeclareMathOperator{\arcsinh}{arcsinh}

\newcommand{\ie}{\textrm{i.e.}\xspace}
\newcommand{\g}{\ensuremath{\boldsymbol{g}}\xspace}
\newcommand{\gE}{\ensuremath{\boldsymbol{\tilde{g}}_{E}}\xspace}
\newcommand{\M}{\ensuremath{\mathpzc{M}}\xspace}

\newcommand{\hm}[2]{\ensuremath{\langle#1,#2\rangle}\xspace}

\def\hksqrt{\mathpalette\DHLhksqrt}
\def\DHLhksqrt#1#2{\setbox0=\hbox{$#1\sqrt{#2\,}$}\dimen0=\ht0
  \advance\dimen0-0.2\ht0
  \setbox2=\hbox{\vrule height\ht0 depth -\dimen0}%
{\box0\lower0.4pt\box2}}
\makeatletter
\def\equalsfill{$\m@th\mathord=\mkern-7mu
  \cleaders\hbox{$\!\mathord=\!$}\hfill
  \mkern-7mu\mathord=$}
\makeatother
\begin{document}
\section{Introduction}\label{sec:intro}
Classical gauge theories are well described through differential geometry, where
a gauge field is represented by a connection on a principal fibre bundle $P$ for
which the structure group is the symmetry group of the theory; see, for example,
\cite{frankel,nakahara,darling,ladies,kerbrat,fulpnorris,mayer,sardanashvily,simpson}
and references therein.

The phenomenon of symmetry breaking also has its own geometric
formulation \cite{ladies,kerbrat,fulpnorris,mayer,sardanashvily,simpson,neeman}: the
reduction of the principal bundle $P$. To this classical picture, the question
that arises is that of how does a quantum field theory select its vacua given by
the different possible reductions of the principal bundle $P$. The canonical
answer to this question is that radiative corrections (beyond the tree-level
approximation) cause the QFT to ``jump'' from the symmetric phase to the
broken-symmetric one(s) \cite{weinbergcoleman}.

However, \cite{weinbergcoleman} already expresses concerns about
some issues (e.g., the last paragraph on page 1894 and its continuation on page
1895) that are more explicitly treated in \cite{garciaguralnik}. For more
examples about these issues, see \cite{borcherds} (pages 27 and 28 discuss
the asymptotic nature of the series expansion, their maximum accuracy and issues
regarding Borel summation) and \cite{fredenhagen}.

Loosely speaking, what \cite{garciaguralnik} does, is to compute the
Schwinger-Dyson equations (for a given QFT) and note that these differential
equations have as many solutions as its order indicates. Moreover, the boundary
conditions that determine each of these solutions yield different possible
values for the parameters of the theory (mass, coupling constants, etc). Thus,
each one of these solutions has its own series expansion (which may or may not
be equivalent to the perturbative series) and particular behavior.

The analysis of the boundary conditions of the Schwinger-Dyson equation being
responsible for the different solutions of the theory, done in \cite{garciaguralnik},
is more in the spirit of the self-adjoint extension of the associated
Hamiltonian, as done in \cite{reedsimon,dunfordschwartz}. Although it may not
seem so at first, \cite{garciaguralnik} is equivalent to a latticized approach to
QFT and, as such, requires 2 types of infinite limits in order to give rise
to its continuum version: the thermodynamic (number of particles in the box) and
the volume (size of the box) limit. These limits do not commute and fiddling
with them (as done in \cite{garciaguralnik}) is analogous to resumming the
perturbative series (as done in \cite{weinbergcoleman}).

The present work has the goal of developing a geometrical method that brings to
the foreground these issues of vacuum structure in QFT, its multitude of
solutions, i.e., its moduli space and the relation between phase structure and
the the topology of the moduli space (different solutions are topologically
inequivalent). The fact that different configuration spaces for distinct
solutions of the equation of motion of a given Lagrangian have different
topologies shows that expansions must be performed separately for each
solution of a theory, \ie, each phase has to be regarded and treated as a
separate theory.

In this sense, this work streamlines how the parameters of a theory (mass,
coupling constants, etc) determine the topology of the vacuum manifold (moduli
space) which, combined with the picture presented in \cite{garciaguralnik},
gives a prescription of how to examine the solution space of a field theory: The
boundary conditions of the Schwinger-Dyson equations determine the parameters of
the theory which, in turn, determine its topology.

In the more modern language of Higgs Bundles (and their associated moduli
spaces, \cite{simpson}), what we are doing is to use an appropriate rescaling of
the metric in such a way as to identify its geodesics with the solutions of the
[local] system in question. In this way, different geodesics characterizes distinct
elements of the moduli space, each one with its own topology, which is
determined by the allowed ranges for the values of the parameters of the
theory. Thus, different phases of the theory are shown to have different
topologies, which brings the subject of topological transitions under a novel
light, where they are represented by phase transitions.

This work is a vast extension of what was done in \cite{symbtopind}, therefore
we decided to start anew. The organization of this work is as follows: in
section \ref{sec:qftjm} we develop the basic tools and apply them to the
classical case; in section \ref{sec:appsqft} we move to quantum field theory and
show several of the properties that have been mentioned above, treating
explicitly the $\lambda\, \phi^4$ example; in section \ref{sec:examples} we use
this technique in gauge theories, revealing the extra structures these theories
have, using the Landau-Ginzburg and Seiberg-Witten models. We conclude with some
remarks about future work.
\section{Classical Field Theory and the Jacobi Metric} \label{sec:qftjm}
A typical action for a scalar field has the form,
\begin{align*}
  S[\phi] &= \int K(\pi,\pi) - V_{\tau}(\phi) \, d^{n}x \; ;\\
  K(\pi,\pi) &= \frac{1}{2}\, \g(\pi,\pi) = \frac{1}{2}\, g_{\mu\,\nu}\,
    \pi^{\mu}\,\pi^{\nu} \; .
\end{align*}
where $K$ is the kinetic quadratic form (bilinear and, in case $\pi\in\mathbb{C}$,
hermitian; defining the inner product \g) and $V_{\tau}$ is the potential, where
the index $\tau$ collectively denotes coupling constants, mass terms, etc.

We want to reparameterize our field using the arc-length parameterization such
that $\boldsymbol{\tilde{\g}}(\tilde{\pi},\tilde{\pi}) = \tilde{g}_{\mu\,
  \nu}\tilde{\pi}^{\mu}\, \tilde{\pi}^{\nu} = 1$, where
$\boldsymbol{\tilde{\g}}$ is the new metric and $\tilde{\pi}^{\mu}$ is the
momentum field redefined in terms of this new parameterization (assuming
$V_{\tau}(\phi)$ contains no derivative couplings). That is, we want to scale our
coordinate system in order to obtain a conformal transformation of the metric
that normalizes our momentum field.

In fact, there are three possible normalizations, depending on the nature
of $\tilde{\pi}$:
\begin{equation*}
  \boldsymbol{\tilde{\g}}(\tilde{\pi},\tilde{\pi}) = 
  \begin{cases}
     1, & \text{spacelike;} \\
     0, & \text{lightlike;} \\
    -1, & \text{timelike.}
  \end{cases}
\end{equation*}

However, this will not concern us here, once our main focus will
be to find the geodesics of $\boldsymbol{\tilde{\g}}$. Therefore, we have a
clear distinction among the three foliations of Lorentzian spaces:

\begin{description}
\item[$\boldsymbol{\tilde{\g}(\tilde{\pi},\tilde{\pi}) > 0}$] represents the
  spacelike-leaf and contains only non-physical objects --- after the arc-length
  reparameterization we will have $\tilde{\g}(\tilde{\pi},\tilde{\pi}) = 1$;
\item[$\boldsymbol{\tilde{\g}(\tilde{\pi},\tilde{\pi}) = 0}$] clearly singular,
  representing the null-leaf where there is no concept of distance, given that
  all objects here are massless;
\item[$\boldsymbol{\tilde{\g}(\tilde{\pi},\tilde{\pi}) < 0}$] represents the
  timelike-leaf and contains the physical objects --- after the arc-length
  reparameterization we will have $\tilde{\g}(\tilde{\pi},\tilde{\pi}) = -1$.
\end{description}

In order to find the arc-length parameterization, we will borrow an idea from
classical mechanics called \emph{Jacobi's metric} \cite{drg}. In the Newtonian
setting, Jacobi's metric gives an intrinsic geometry for the configuration
space (resp. phase space), where dynamical orbits become geodesics, i.e., it
maps every Hamiltonian flow into a geodesic one; therefore, solving the
equations of motion implies finding the geodesics of Jacobi's metric and
vice-versa. This can be done for any closed non-dissipative system (with total
energy $E$), regardless of the number of degrees of freedom.

Before we proceed any further, let us take a look at a couple of simple examples
in order to motivate our coming definition of Jacobi's metric.

\paragraph{Harmonic Oscillator} \label{par:sho}
Let us consider a model similar to the harmonic oscillator (in
$(1+0)$-dimensions): without further considerations, we simply allow for the
analytic continuation of the frequency: $\mu = +\omega^2$ or $\mu =
-\omega^2$. Its Lagrangian is given by $L = \tfrac{1}{2}\, \dot{q}^2 \mp
\tfrac{\mu}{2}\, q^2$, where $\mu > 0$, from which we conclude that for a fixed
$E$ such that $E = \tfrac{1}{2}\, (\dot{q}^2 \pm \mu\,q^2)$, we have $\dot{q} =
\tfrac{dq}{dt} = \hksqrt{2\, (E \mp \tfrac{\mu}{2}\, q^2)}$.

So, in the spirit of what was said above, we want to find a reparameterization
of the time coordinate in order to have the normalization $\dot{q} = 1$, where
the dot represents a derivative with respect to this new time variable.

Thus,

\begin{align*}
  \frac{dq}{dt} &= \frac{dq}{ds}\, \frac{ds}{dt} = \hksqrt{2\, \biggl(E \mp
    \frac{\mu}{2}\, q^2\biggr)} \; ;\\
  \text{if}\quad \frac{ds}{dt} &= \hksqrt{2\, \biggl(E \mp \frac{\mu}{2}\,
    q^2\biggr)} \Rightarrow \frac{dq}{ds} \equiv 1\;; \\
  \therefore\; ds &= \hksqrt{2\, \biggl(E \mp \frac{\mu}{2}\, q^2\biggr)}\, dt\; ; \\
  \Rightarrow\; \gE &= 2\, \biggl(E \mp \frac{\mu}{2}\, q^2\biggr)\, \g \; .
\end{align*}

As expected, the new metric, \gE, is a conformal transformation of the original
one, \g; and under this new metric, our original Lagrangian is simply written
as,

\begin{align*}
  L(q,\dot{q}) &= \frac{1}{2}\, \dot{q}^2 \mp \frac{\mu}{2}\, q^2 \; ;\\
  &= \frac{1}{2}\, g_{i\, j} \dot{q}^i\,\dot{q}^j \mp\frac{\mu}{2}\, q^2\; ;\\
  &= \frac{1}{2}\, \g(\dot{q},\dot{q}) \mp\frac{\mu}{2}\, q^2\; ;\\
  \therefore\; L(q,\dot{q}) &= \gE(\dot{q},\dot{q}) \; .
\end{align*}

Now, as we said above \cite{drg}, the Euler-Lagrange equations for this
conformally transformed Lagrangian are simply the geodesics of the metric
\gE. The task of finding the geodesics $\gamma(s)$ of the \gE metric can be more
easily accomplished with the help of the normalization condition,
$\gE\bigl(\tfrac{d\gamma}{ds},\tfrac{d\gamma}{ds}\bigr) = 1$, and the initial
condition $\gamma(s=0) = 0$, given that, in this fashion, we only need to solve
a first order differential equation:

\begin{align*}
  \gE\biggl(\frac{d\gamma}{ds}, \frac{d\gamma}{ds}\biggr) &= 1 \; ;\\
  \Rightarrow\; 2\, \biggl(E \mp \frac{\mu}{2}\, \gamma^2\biggr)\,
    \g\biggl(\frac{d\gamma}{ds}, \frac{d\gamma}{ds}\biggr) &= 1 \; ;\\
  \therefore\; \gamma'(s) = \frac{d\gamma}{ds} &= \hksqrt{\frac{1}{2\,\bigl(E
      \mp \frac{\mu}{2}\, \gamma^2\bigr)}} \; ;\\
  \text{with the initial condition:}\quad \gamma(0) &= 0 \; .
\end{align*}

Here we should note that solving for $\gE(\gamma',\gamma') = 1$ (spacelike-leaf)
is analogous to solving $\gE(\gamma',\gamma') = -1$ (timelike-leaf), once the
two cases are symmetrical about the origin.

Analytically solving the equation above yields 2 possible answers, depending on
the particular form of the potential ($\mu > 0$ in both cases):

\begin{enumerate}
\item $V_{+} = +\mu\, \gamma^{2}/2$: For the case of a positive pre-factor, we
  get that $\gamma\, \hksqrt{\mu\, (2\, E - \mu\, \gamma^{2})} + 2\, E\,
  \arcsin\bigl(\gamma\, \hksqrt{\mu/2E}\bigr) - 2\, s\, \hksqrt{\mu} = 0$; and
\item $V_{-} = -\mu\, \gamma^{2}/2$:  For the case of a negative pre-factor, we
  find that $\gamma\, \hksqrt{\mu\, (2\, E + \mu\, \gamma^{2})} + 2\, E\,
  \arcsinh\bigl(\gamma\, \hksqrt{\mu/2E}\bigr) - 2\, s\, \hksqrt{\mu} = 0$.
\end{enumerate}

It is clear from the expression for $V_{+}$ that $\gamma^{2} \leqslant 2E/\mu$,
i.e., the length of the [classical] geodesic is bounded; this does not happen
with $V_{-}$.

These geodesics, $\gamma_{\pm}$, clearly depend on the parameters $E$ and $\mu$;
therefore, in order to plot $\gamma(s)$, we have to make 2 distinct choices:
$E/\mu = 1$ (left plot) and $E/\mu = -1$ (right plot). The last plot
comparatively depicts both geodesics.

\vspace{.7cm}
\begin{center}
  \includegraphics[scale=0.5]{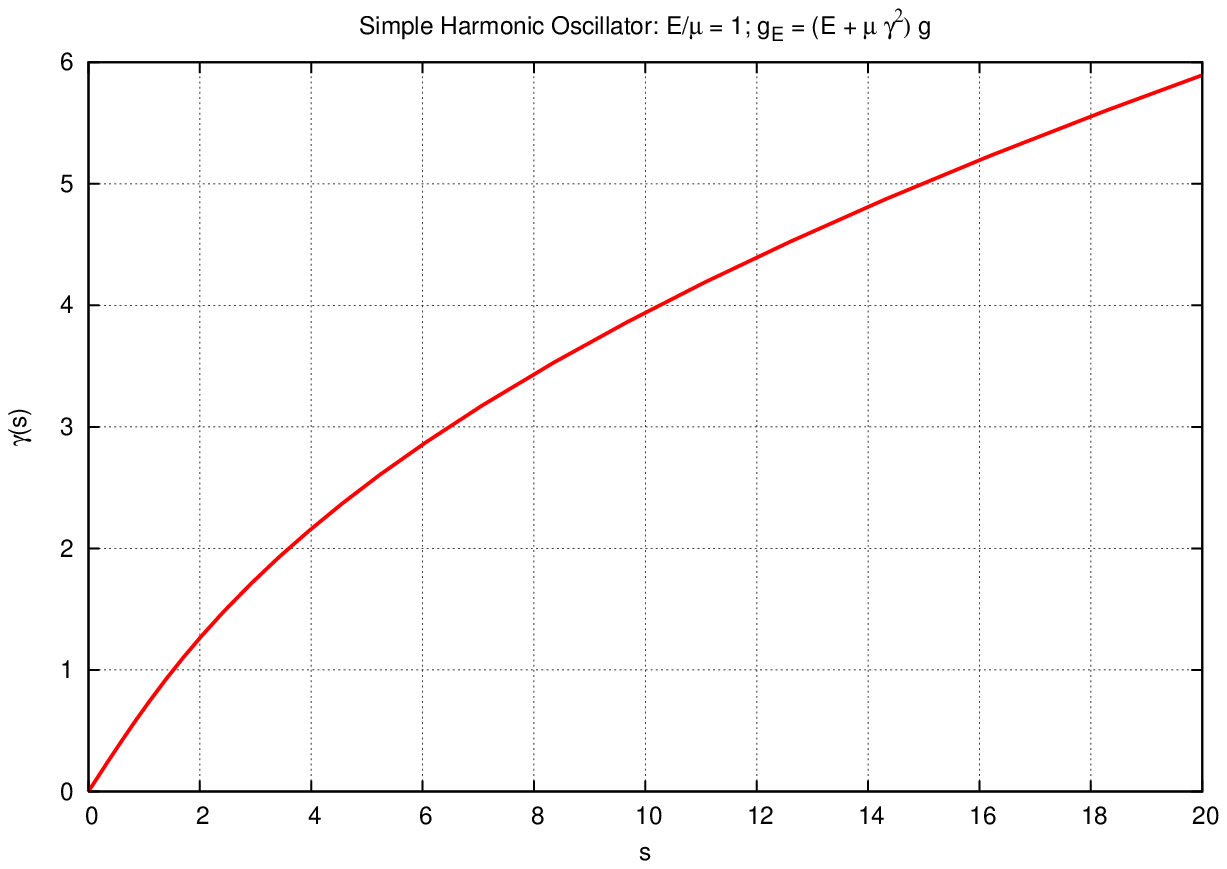} \hfill
  \includegraphics[scale=0.5]{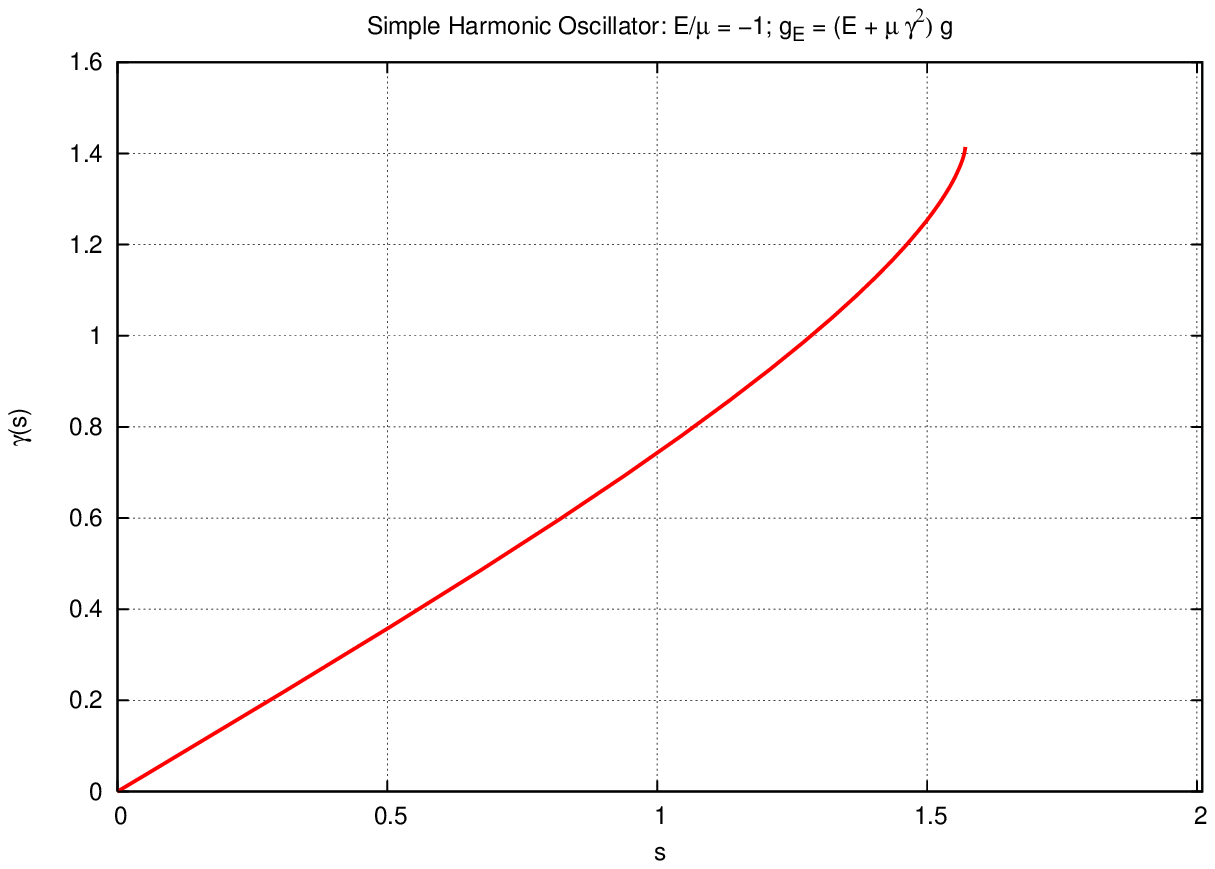} \hfill\\
  \includegraphics[scale=0.5]{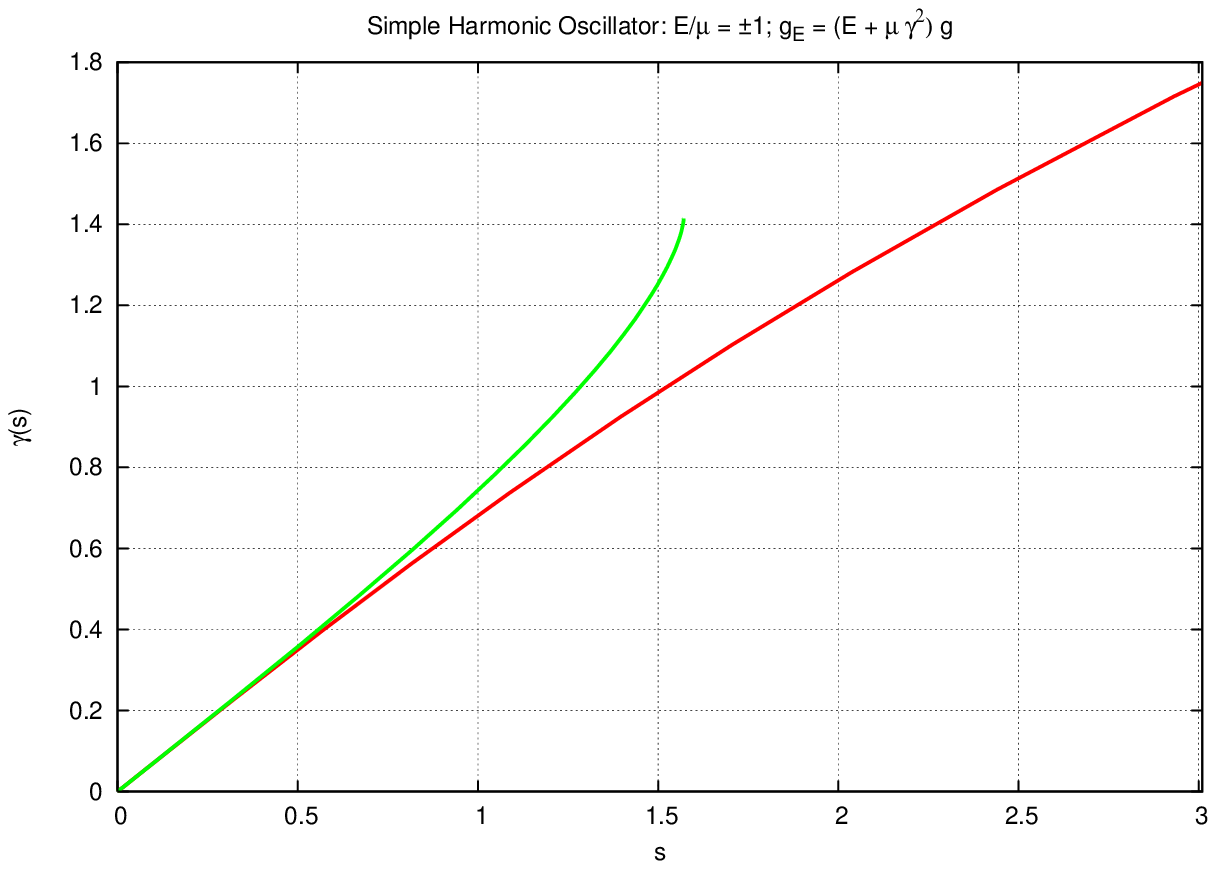}
  \bigskip \bigskip \bigskip

  {\footnotesize \noindent\textbf{Figure 1:} The top-left-corner plot shows the
    geodesic for $E/\mu = 1$, while the top-right-corner one depicts it for
    $E/\mu = -1$. The bottom-center plot superimposes both of them.}
\end{center}

This ``harmonic oscillator'' model already portrays the features that we want to
identify in forthcoming applications of this technique: changing the values of
the parameters of the potential we can identify 2 distinct types of geodesics
which will be related to different solutions on the coming examples.

Note, however, that in general the parameters will not be able to vary freely:
they will belong to some specified set; and they will be related to each other,
i.e., we will be able to find a function of the parameters that constrains their
behavior --- this will establish ``dualities'' among the parameters of a given
theory, as will be shown below.

\paragraph{Cubic Potential} \label{par:airy}
For completeness sakes, let us try another example, that of a cubic potential
given by: $L = \tfrac{1}{2}\, \g(\dot{q},\dot{q}) + q^3/3$. It's Jacobi Metric
is given by $\gE = 2\, (E + q^3/3)\, \g$ and the equation we need to solve in
order to find the geodesics, $\gamma(s)$, of this metric is,

\begin{align*}
  \gE(\gamma',\gamma') = 2\, (E + q^3/3)\, \g(\gamma',\gamma') &= 1 \; ;\\
  2\, (E + q^3/3)\, \bigl(\gamma'(s)\bigr)^2 &= 1 \; ; \\
  \text{with the initial condition:}\quad \gamma(0) &= 0 \; .
\end{align*}

Once again, we should note that solving for $\gE(\gamma',\gamma') = -1$
(timelike-leaf) is analogous to what is being done above: the graphs are
reflected with respect to each other.

This example is slightly different from the previous one for the following
reason: before, it was the relative values of $E$ and $\mu$ that determined our
two solutions, i.e., one for $E/\mu \geqslant 0$ and one for $E/\mu < 0$. Now,
this cubic theory has \emph{no} free parameters in its potential, therefore the
geodesics are labeled by the arbitrary parameter $E$.

Just as before, the metric will be [artificially] degenerate when $E =
V(\gamma)$, which tells us that $\gamma \geqslant (-3\, E)^{1/3}$: when $E
\geqslant 0$ the geodesic is allowed to have any length, but when $E < 0$ the
geodesic vanishes for some time, after which it starts to grow.

This is very interesting because it essentially says that there is a ``lag
time'' before this solution comes alive: this leaf is non-existent for some
``proper time'' and then it springs into being quite abruptly.

\begin{center}
  \includegraphics[scale=0.5]{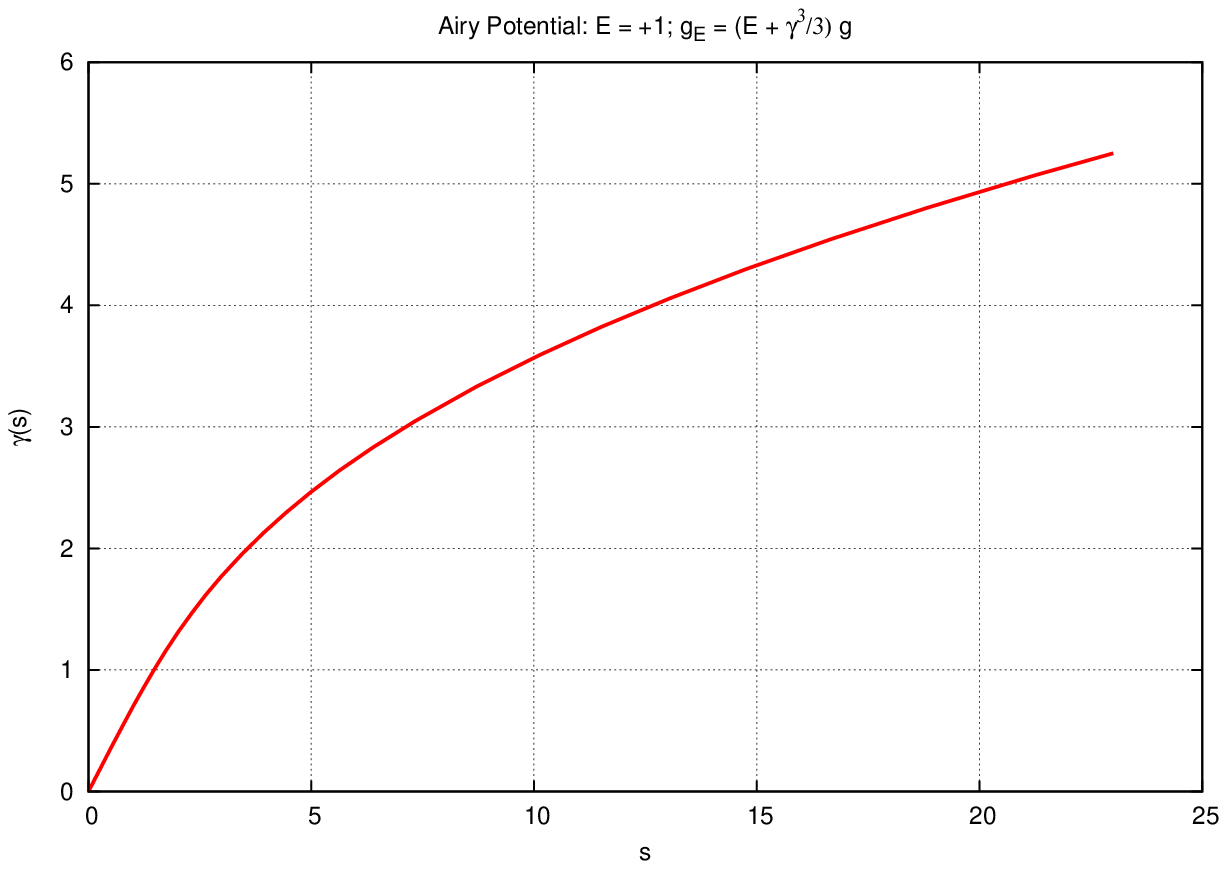} \hfill
  \includegraphics[scale=0.5]{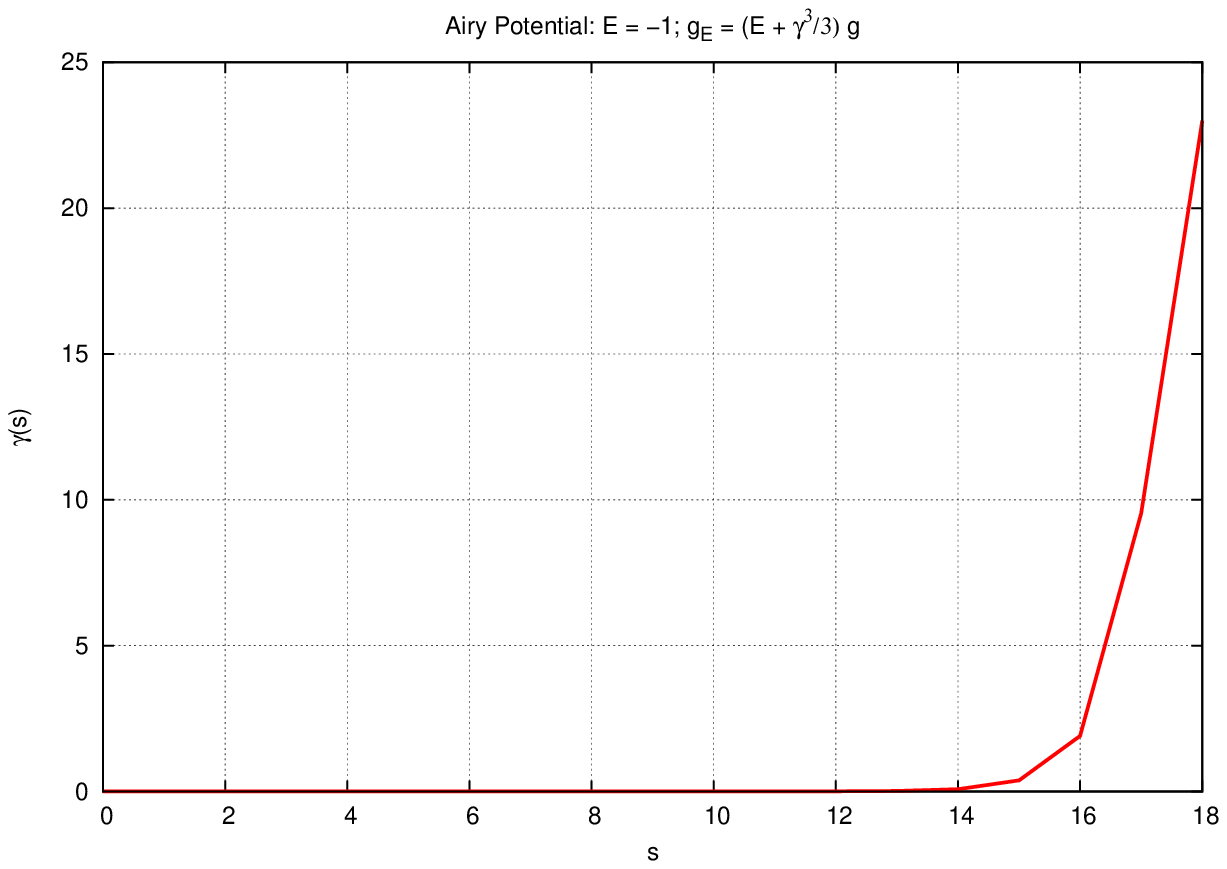} \hfill\\
  \includegraphics[scale=0.5]{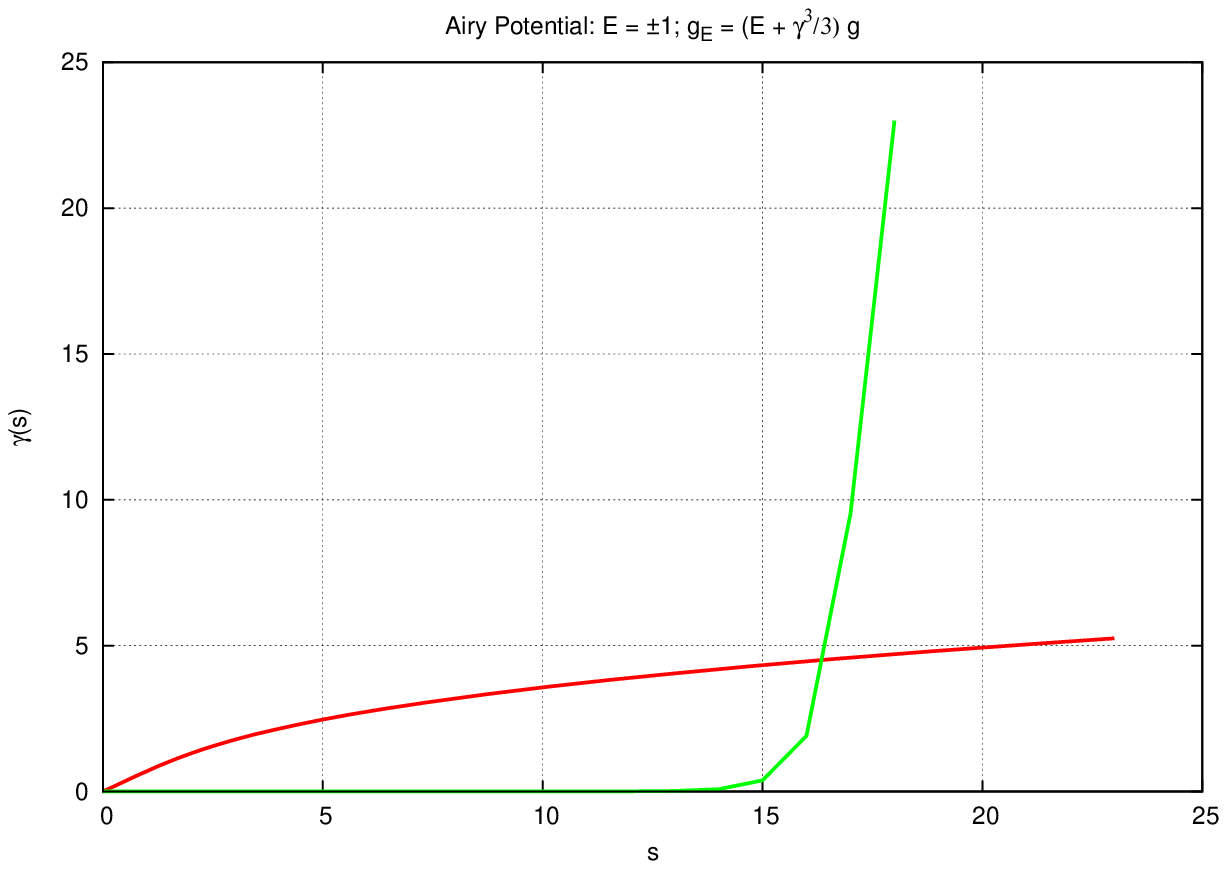}
  \bigskip \bigskip \bigskip

  {\footnotesize \noindent\textbf{Figure 2:} The top-left-corner plot shows the
    geodesic for $E = 1$, while the top-right-corner one depicts it for
    $E = -1$. The bottom-center plot superimposes both of them.}
\end{center}

Now that we are done with our examples, we are ready to find the generalization
of Jacobi's metric to the field theoretical setting can be accomplished via the
definition of the following conformally transformed metric:

\begin{align}
  \nonumber
  L &= \frac{1}{2}\, \g(\pi,\pi) - V_{\tau}(\phi) \; ; \\
  \nonumber
  &\equiv \gE^{\tau}(\tilde{\pi},\tilde{\pi}) \; ;\\
  \intertext{where,}
  \label{eq:jacobimetric}
  \gE^{\tau} &= 2\, \bigl(E - V_{\tau}(\phi)\bigr)\, \g \; ;
\end{align}
and $E$ (an arbitrary parameter) is the total energy of the system.
\section{Applications in Quantum Field Theory} \label{sec:appsqft}
We will use the approach in terms of Feynman Path Integrals in order to make
things more straightforward. However, we could as well talk in terms of the
phase space of a given QFT and its vacuum manifold, i.e., its moduli space.

Starting from the partition function, we have the following ($\hbar = 1$):

\begin{align*}
  \mathcal{Z}[J] &= \mathcal{N}\, \varint e^{i\, S[\phi] + i\int J(x)\,
    \phi(x)\, d^dx}\, \mathcal{D}\phi \; ;\\
  &= \mathcal{N}\, \varint e^{i\, \int \gE^{\tau}(\tilde{\pi},\tilde{\pi}) + J(x)\,
    \phi(x)\, d^dx}\, \mathcal{D}\phi \; ;
\end{align*}
where $\mathcal{N}$ is a normalization constant such that $\mathcal{Z}[J=0] =
1$, and we have already written the Action in terms of Jacobi's metric.

Now, let us expand the partition function above in terms of its classical part
and its quantum fluctuations, i.e., $\phi = \phi_{\text{cl}} + \delta\phi$,
and the classical Action is given by $S_{\text{cl}} = S[\phi_{\text{cl}}] =
\int_{\mathpzc{M}} \gE^{\tau}(\tilde{\pi}_{\text{cl}},\tilde{\pi}_{\text{cl}})$, where
$\mathpzc{M}$ is the particular region of spacetime where the integration is
performed and $\pi = \ed\phi$ --- for this calculation we are assuming the
fluctuations vanish at the boundary, $\delta\phi|_{\partial\mathpzc{M}} =
0$. Thus, we have that,

\begin{align}
  \nonumber
  \mathcal{Z} &= \varint \exp\bigl\{i\, S[\phi_{\text{cl}} +
    \delta\phi]\bigr\}\, \mathcal{D}\phi \; ; \\
  \label{eq:clfldecomp1}
  &= \varint \exp\biggl\{i\, \int_{\mathpzc{M}}
    \gE^{\tau}(\tilde{\pi}_{\text{cl}},\tilde{\pi}_{\text{cl}}) +
    \gE^{\tau}(\delta\tilde{\pi},\delta\tilde{\pi})\biggr\} \, \mathcal{D}\phi \; ; \\
  \label{eq:clfldecomp2}
  &= e^{i\, S_{\text{cl}}}\, F_{\tau}[\partial\mathpzc{M}] \; ; \\
  \intertext{where,}
  \label{eq:quantumfl}
  F_{\tau}[\partial\mathpzc{M}] = \varint e^{i\, S[\delta\phi]}\,
    \mathcal{D}(\delta\phi) &= \varint e^{i\, \int_{\mathpzc{M}}
      \gE^{\tau}(\delta\tilde{\pi},\delta\tilde{\pi})}\,
    \mathcal{D}(\delta\phi) \approx \frac{1}{\hksqrt{\det\bigl(-D^2_{\tau}\bigr)}} \; ;
\end{align}
where $F_{\tau}[\partial\mathpzc{M}]$ is the [quantum] fluctuation part of the
partition function, the index $\tau$ denoting collectively the parameters of the
potential (mass, coupling constants, etc), and $D^2_{\tau} = \nabla^{\mu}\,
\nabla_{\mu}$ is the Laplace-Beltrami operator constructed from the covariant
derivative $(\nabla_{\mu})$ associated to the Levi-Civita connection of
$\gE^{\tau}$. There are two important things to note from \eqref{eq:quantumfl}:
the fluctuation term can only depend on the parameters of the potential and on
the boundary $\partial\mathpzc{M}$, i.e., surface terms (that we assumed
vanishing); and its structure is analogous to that of the classical part, in
that $\gE^{\tau}$ is the same on both.

However, the above presumes that there is only one solution to the theory in
question. But, the method we are developing is exactly to use Jacobi's metric in
order to find and classify \emph{all} of the solutions of the theory given,
i.e., we want to be able to use this tool to study the moduli space of the
problem at hand. Therefore, we need to generalize the situation above for the
case of \emph{many} solutions, which is not a difficult task:

\begin{align}
  \label{eq:multiclfldecomp1}
  \mathcal{Z} &= \prod_{\nu=1}^{N} e^{i\, S_{\text{cl}}^{\nu}}\,
    F_{\tau}^{\nu}[\partial\mathpzc{M}] \; ;\\
  \label{eq:multiclfldecomp2}
  &\approx \prod_{\nu=1}^{N} \frac{e^{i\,
      S_{\text{cl}}^{\nu}}}{\hksqrt{\det\bigl(-D^2_{\tau}\bigr)}} \; ;
\end{align}
where $\nu$ counts the number of different solutions (i.e., the number of
critical points of $S[\phi]$, including its multiplicity) denoted by
$\phi^{\nu}_{\text{cl}}$, and $S_{\text{cl}}^{\nu} =
S[\phi^{\nu}_{\text{cl}}]$. Note that, in order to find all of the possible
critical points of $S[\phi]$, we need to take into account its
$\tau$-dependence, and in doing so we are implicitly assuming that $\tau \in
\mathbb{C}$, ensuring we are able to find them all.

There are two major observations to be done at this point:
\begin{enumerate}
\item There are two kinds of discontinuities present in the above construction,
  \cite{berrystokes}: first, critical \emph{points} $\phi^{\nu}_{\text{cl}}$ can
  coalesce, which happens in the complexified catastrophe set in parameter
  space; second, critical \emph{values} $\Im(i\, S[\phi^{\nu}_{\text{cl}}]) =
  S^{\nu}_{\text{cl}}$ can coalesce, which happens on the Stokes set in
  parameter space and corresponds to the appearance or disappearance of a
  subdominant exponential in a ``non-local bifurcation''. These correspond,
  respectively, to realizing that the partition function is to be taken over
  \emph{complex} fields (rather than real ones, as is customary), where the contour of
  integration (rendering the partition function finite, \cite{garciaguralnik}) will ultimately
  determine the parameter space (these are the Lee-Yang zeros of our theory);
  and the coalescing of $S^{\nu}_{\text{cl}}$ represents the Stokes phenomena of our
  theory. Both of these will determine the phase structure of our problem (see
  \cite{garciaguralnik}). Note that it is the contour of integration that
  connects both of these, once the appropriately chosen range of $\phi$ will
  render $e^{i\, S[\phi]}$ convergent.
\item The asymptotic behavior of the partition function depends \emph{only} on
  the critical points of $S$, i.e., on $\phi^{\nu}_{\text{cl}}$. Thus, when
  $S[\phi]$ (resp. $V[\phi]$) is a Morse function, i.e, smooth with no degenerate
  critical points, we can use Morse's lemma to show that the critical points are
  isolated, keeping in mind that the number of isolated points is a topological
  invariant. (It is worth noting this can be generalized in the thermodynamical
  limit via the Morse-Palais lemma, just as we can relax the condition of
  non-degeneracy of the critical points via Morse-Bott theory.) Using this, it
  can be shown that $\phi$-space (resp. Phase Space) is a CW-complex with a $\nu$-cell
  for each critical point of index $\nu$: the fluctuation term contains the
  Maslov-Morse index (in $\gE^{\tau}$ by means of its dependence on
  $V_{\tau}[\phi]$) that accounts for the discretization of Path Space and
  corrects for the thermodynamic limit of the particular solution in question,
  i.e., every time the denominator in \eqref{eq:multiclfldecomp2} vanishes,
  $F_{\tau}^{\nu}[\partial\mathpzc{M}]$ passes a singularity in such a way as to
  ensure the proper phase; the phase factor that arises in this way is nothing
  but $e^{i\, \alpha\, \nu}$, where $\alpha$ is some angle and $\nu$ counts the
  number of zeros (with multiplicity) encountered along the particular path in
  question --- it is called the Maslov-Morse index.
\end{enumerate}

Now, we are ready to employ this machinery in order to draw several important
conclusions. Here are them:

\begin{enumerate}
\item The partition function, seen as a function of the parameters of the
  potential, $\mathcal{Z} = \mathcal{Z}[\tau] = \mathcal{Z}[\text{mass},
  \text{coupling constants}]$, is a \emph{meromorphic} function: it is
  holomorphic on a subset of $\mathbb{C}$ except for the set of isolated points
  given by the values of the parameters along Stokes' lines (resp. critical
  lines of phase transition). We have to keep in mind that the parameter space
  had to be complexified in order to yield all possible solutions to the theory
  given; another way to think about this is in terms of the contours that render
  the partition function finite: changing these contours (looking for all
  possible ones that make the partition function converge) will affect the
  allowed values for the parameters, i.e., these contours ultimately determine
  the parameter space, as done in \cite{garciaguralnik} (see also
  \cite{mollifier}).
\item Under the [full] elliptic modular group, $\Gamma =
  \mathrm{SL}(2,\mathbb{Z}) = \bigl\{ \bigl(
  \begin{smallmatrix}
    a & b\\
    c & d
  \end{smallmatrix}
  \bigr) \;|\; a=1,\, b=0,\, c=0,\, d=1\bigr\}$ (see \cite{mfmf}), the partition
  function is a \emph{modular function}, i.e., for any $\mathds{M} \in \Gamma$
  we have that $\mathcal{Z}[\mathds{M}\, \tau] = \mathcal{Z}[\tau]$, where
  $\mathds{M}\, \tau = \tfrac{a\, \tau + b}{c\, \tau + d}$. (For a fuller
  appreciation of the importance of this fact, see, e.g., \cite{3dgravity}.)
\item The Action, written in terms of Jacobi's
  metric, $S[\phi] = \int_{\mathpzc{M}} \gE(\tilde{\pi},\tilde{\pi})$, can be
  thought of as the Morse-theoretic Energy functional, $E[\gamma] =
  \int\g(\gamma',\gamma')$, for the path $\gamma$, where $\gamma' =
  d\gamma/ds$. Therefore, we can readily say that all of the critical points of
  $S[\phi]$ are given by minimal [Lagrangian] manifolds, and, just like $E''$,
  $S''=\delta^2 S/\delta\phi(x)\delta\phi(y)$ is a well defined symmetric bilinear
  functional. This implies that $S'' = 0$ if, and only if, $\tilde{\pi}$ is a
  Jacobi field, which, in turn, implies that $\tilde{\pi}|_{\partial\mathpzc{M}} =
  0$. Therefore, using the split of the partition function
  \eqref{eq:multiclfldecomp1} in terms of a classical part and its quantum
  fluctuations, we see that the vacuum manifold (moduli space) of the quantum
  theory is given by its classical minimal Lagrangian manifold with the quantum
  corrections being given by extensions of it via Jacobi fields. That is, the
  classical minimal manifold is extended via the gluing of the quantum
  fluctuation manifold obtained as a solution to the equations of motion coming
  from $S_{\text{fl}} = \int_{\mathpzc{M}}
  \gE^{\tau}(\delta\tilde{\pi},\delta\tilde{\pi})$. Thus, the quantum
  corrections are handles attached to the classical solution, without changing
  the classical topology.
\item Continuing the reasoning above, it is not difficult to see that a phase
  transition will happen when the quantum corrections (in terms of handle
  attachments) actually do change the topology of the particular solution in
  question. In fact, if $V_{\tau}[\phi]$ crosses a critical point of index
  $\nu$, the handle to be attached is a $\nu$-cell (i.e., a $\nu$-simplex) ---
  the connection with what has been said before should be
  straightforward. Analogously, we can say the following: let $\Sigma_{\tau}^E = 
  V^{-1}_{\tau}[E] = \{ \phi \;|\; V_{\tau}[\phi] = E\}$, i.e.,
  $\Sigma_{\tau}^E$ is the set of all field configurations which have the same
  potential energy --- this manifold is the configuration space (resp. moduli
  space), an equipotential surface. Thus, the family of equipotentials
  $\{\Sigma_{\tau}^E\}_{E\in \mathbb{R}}$ foliates the configuration space
  (moduli space) in such a way that if $\Sigma_{\tau}^E$ is
  $\mathscr{C}^{\infty}$-diffeomorphic to $\Sigma_{\tau}^{\bar{E}}$ (for two
  different values of the energy, $E$ and $\bar{E}$) then there is no phase
  transition. Conversely, a phase transition will be characterized by the
  existence of a certain critical value $E_c$ such that
  $\{\Sigma_{\tau}^{E}\}_{E < E_c}$ is \emph{not}
  $\mathscr{C}^{\infty}$-diffeomorphic to
  $\{\Sigma_{\bar{\tau}}^{\bar{E}}\}_{\bar{E} > E_c}$ (note that, upon a phase
  transition, the parameters of the potential change from $\tau$ to
  $\bar{\tau}$). Further, the topological difference between these two leaves is
  a $\nu$-cell, where $\nu$ is the index of the critical point $E_c$. Loosely
  speaking, it can be said that the origin of phase transitions is this change
  in topology. Therefore, different phases of the theory are topologically
  inequivalent.
\item Finally, let us note that all of these conclusions we have drawn so far
  have one last implication, a topological constraint: the Euler Characteristic,
  $\chi$, computed from \gE gives the quantization rules [for the
  energy]. Therefore, our theory can be topologically quantized; in fact, each
  topologically inequivalent leaf $\{\Sigma_{\tau}^{E}\}$ has its own
  $\chi_{\tau}^{E}$ and, thus, its own quantization rules.
\end{enumerate}

With all of these facts in hand, let us see what is the framework they imply:
given a theory, $L = \g(\pi,\pi) + V_{\tau}[\phi]$, we can readily construct its
Jacobi metric, $\gE^{\tau} = 2\, (E - V_{\tau})\, \g$: from this, we can do two
things, either compute the geodesics of $\gE^{\tau}$ and classify their Jacobi
fields in terms of $\tau$, or calculate its Euler Characteristics,
$\chi_{\tau}^{E}$, and find the quantization rules (bearing in mind that they
will vary with $\tau$). The different geodesics (corresponding to different
Jacobi fields) will label diffeomorphically equivalent foliations
$\{\Sigma_{\tau}^{E}\}$, while the order of the zeros of
$F_{\tau}^{\nu}[\partial\mathpzc{M}]$ will correspond to the $\nu$-cells associated
with a particular phase transition --- furthermore, they will be responsible for
the Lee-Yang zeros and the Stokes phenomena of the theory. In turn, this gives
the partition function a meromorphic and a modular character when seen with
respect to $\tau$.

All of this was possible because we extended the parameter space, allowing
$\tau$ to be complex. This is completely analogous to complexifying the solution
space (moduli space) of the theory given, a fact which is made clear in
\cite{garciaguralnik} (and employed in lattice calculations in
\cite{mollifier}).

Lastly, using $\tau$, it will be possible to construct ``dualities'' between
different phases. However, in general, these dualities will be non-trivial
combinations of the parameters of the theory (as opposed to what we found in the
simple examples above, where $E/\mu \mapsto -E/\mu$ or $E \mapsto -E$ did the
job).
\subsection{The $\boldsymbol{\lambda\, \phi}^{\mathbf{4}}$ Potential}\label{subsec:lp4}
This theory is defined for scalar-valued fields, $\phi$, by the Lagrangian $L =
\tfrac{1}{2}\, \bigl(\g(\pi,\pi) - \mu\, \phi^2 - \tfrac{\lambda}{2}\, 
\phi^4\bigr)$, i.e., the potential is given by $V(\phi) = \tfrac{1}{2}\,
\bigl(\mu\, \phi^2 + \tfrac{\lambda}{2}\, \phi^4\bigr)$; which is invariant by
$\mathbb{Z}_2$-reflection: $\phi \mapsto -\phi$.

The Jacobi metric for this theory is given by $\gE^{\tau} = 2\, \bigr(E -
\tfrac{\mu}{2}\, \phi^2 - \tfrac{\lambda}{4}\, \phi^4\bigr)\, \g$. And
solving the normalization condition $\gE^{\tau}(\gamma',\gamma') = 1$ --- where
$\gamma(s)$ is the geodesic we want to compute and $\gamma' =
\tfrac{d\gamma}{ds}$, where $s$ is the arc-length parameter --- should give us
the geodesic structure of the vacuum manifold (resp. moduli space):

\begin{align}
  \nonumber
  \gE^{\tau}(\gamma',\gamma') &= 1 \; ;\\
  \label{eq:lp4geq}
  2\, \bigr(E - \tfrac{\mu}{2}\, \gamma^2 - \tfrac{\lambda}{4}\,
    \gamma^4\bigr)\, (\gamma')^2 &= 1 \; ; \\
  \nonumber
  \hksqrt{E - \tfrac{\mu}{2}\, \gamma^2 - \tfrac{\lambda}{4}\,
    \gamma^4}\, d\gamma &= \tfrac{1}{2}\, ds \; .
\end{align}

The full picture presents itself upon a more detailed analysis of
$\int\hksqrt{E - \mu\, \gamma^{2}/2 - \lambda\, \gamma^{4}/4}\, d\gamma$, which
is the [elliptic] integral that needs to be solved in order to find
$\gamma(s)$. Therefore, it is useful to consider the polynomial $P(\gamma) =
(\gamma^{2} - r_1)\, (\gamma^{2} - r_2)$, where $(-\lambda/4)\, P(\gamma) = E -
\mu\, \gamma^{2}/2 - \lambda\, \gamma^{4}/4$ and $r_{1,2} = -\bigl(\mu \pm
\hksqrt{4\, E\, \lambda + \mu^{2}}\bigr)/\lambda$. It is the \emph{discriminant}
of this polynomial $P(\gamma)$ that will, ultimately, determine the different
solutions of the theory: $\Delta = (r_1 - r_2)^{2} = \lambda\, E + \mu^{2}/4
\lesseqgtr 0$.

As mentioned above, the ``dualities'' between different phases involves
non-trivial combinations of the parameters of the theory. In this case, the
dualities are given by the possible values of the discriminant above: $\Delta >
0$, $\Delta = 0$ or $\Delta < 0$.

The two inequalities, $\Delta > 0$ and $\Delta < 0$, are related by the
reflection of the $\tau$ parameter of this theory, where $\tau = \mu^2/\lambda$,
i.e., by the analytic continuation of the mass parameter such that $\mu^2
\mapsto -\mu^2$, which implies that $\tau \mapsto -\tau$ --- note that this can
be obtained by a modular transformation, $\mathds{M}\, \tau = \tfrac{a\, \tau +
  b}{c\, \tau + d} = -\tau$, such that $b=0=c$, $d=1$ and $a=-1$; further, the
analytic continuation of $\mu$ is such that $\mu \mapsto \pm i\, \mu$, or more
generically as $\mu \mapsto e^{\pm i\, \pi\, \nu/2}\, \mu$, for odd $\nu$ (this is
related to the Stokes phenomena discussed previously, where $\nu$ selects a
certain Riemann sheet for this analytic continuation, which depends on the
Maslov-Morse index $\nu$) ---; meanwhile, the case $\Delta = 0$ has to be
computed separately, once it establishes a fixed value $\tau = -4\, E$ --- note
that when $\Delta = 0$ the \emph{resultant} of $P(\gamma)$ and its derivative,
$P'(\gamma)$, also vanishes, this resultant being given by $\Res(P,P') =
-\tfrac{\lambda}{4}\, \Delta$, which says that $P$ and $P'$ have a common root
(which happens when $\tau = \tfrac{\mu^2}{\lambda} = -4\, E$).

This whole framework is potentially a very interesting result, once
discriminants are closely related and contain information about
\emph{ramifications} (``branching out'' or ``branches coming together'') in
Number Theory. As we just saw, these ramifications are related to the
topological structure of each solution, where the Maslov-Morse index, $\nu$,
selects the appropriate Riemann sheet (for the analytic continuation) and the
topology of the problem (determined by the attachment of a
$\nu$-cell). Therefore, this duality relating the two different solutions, for
$\tau$ and $-\tau$, is given by a [particular] modular transformation of $\tau$
(resp. analytic continuation of $\mu$), which turns out to be a measure of the
ramifications of the problem.

The analytical answers for the geodesic $\gamma(s)$ are found to be given by the
following implicit equations:

\begin{description}
\item[$\boldsymbol{\Delta = 0 \, ,\; \Res(P,P') = 0:}$]
  \begin{equation}
    \label{eq:deltaEQ0}
    \frac{1}{3}\, \gamma^{3} + \frac{\mu}{\lambda}\, \gamma + s = 0\; ;
  \end{equation}
\item[$\boldsymbol{\Delta > 0 \, ,\; \Res(P,P') < 0:}$]
  \begin{equation}
    \label{eq:deltaGT0}
    \begin{split}
      3\, s &+ \gamma\, \hksqrt{(\gamma^{2} - r_1)\,(\gamma^{2} - r_2)} + 2\, r_1\,
      \hksqrt{r_2}\,
      F\biggl(\frac{\gamma}{\hksqrt{r_1}}\boldsymbol{;}
      \hksqrt{\tfrac{r_1}{r_2}}\biggr) -\\
      &- \frac{2}{3}\, \frac{\mu}{\lambda}\,
      \hksqrt{r_2}\, \Biggl[F\biggl(\frac{\gamma}{\hksqrt{r_1}}\boldsymbol{;} 
      \hksqrt{\tfrac{r_1}{r_2}}\biggr) - E
      \biggl(\frac{\gamma}{\hksqrt{r_1}}\boldsymbol{;}
      \hksqrt{\tfrac{r_1}{r_2}}\biggr)\Biggr] = 0 \; ;
    \end{split}
  \end{equation}
\item[$\boldsymbol{\Delta < 0 \, ,\; \Res(P,P') > 0:}$]
  \begin{equation}
    \label{eq:deltaLT0}
    \begin{split}
      3\, s &+ \gamma\, \hksqrt{(\gamma^{2} - r_1)\,(\gamma^{2} - r_2)} + 2\, r_1\,
      \hksqrt{r_2}\,
      F\biggl(\frac{\gamma}{\hksqrt{r_1}}\boldsymbol{;}
      \hksqrt{\tfrac{r_1}{r_2}}\biggr) -\\
      &- \frac{2}{3}\, \frac{\mu}{\lambda}\,
      \hksqrt{r_2}\, \Biggl[F\biggl(\frac{\gamma}{\hksqrt{r_1}}\boldsymbol{;} 
      \hksqrt{\tfrac{r_1}{r_2}}\biggr) - E
      \biggl(\frac{\gamma}{\hksqrt{r_1}}\boldsymbol{;}
      \hksqrt{\tfrac{r_1}{r_2}}\biggr)\Biggr] = 0 \; .
    \end{split}
  \end{equation}
\end{description}

Note that, in the equations above, $F(z;k)$ is the incomplete elliptic integral
of the first kind, while $E(z;k)$ is the incomplete elliptic integral of the
second kind. Moreover, $r_{1,2}$ are defined (as shown above) as the solutions to
the polynomial equation $P(\gamma) = 0$: $r_{1,2} = -\bigl(\mu \pm \hksqrt{\mu^{2} +
  4\, \lambda \, E}\bigr)/\lambda$. On top of this, although \eqref{eq:deltaGT0}
and \eqref{eq:deltaLT0} have the same form, they will yield distinct solutions,
once the relation given by $\Delta = \lambda\, E + \mu^{2}/4$ will either be
positive or negative, which, in turn, affects the outcome of $r_{1,2}$ --- as
already discussed above for $\tau \mapsto -\tau$.

The graphical results are shown below, where $\tfrac{E}{\lambda} = 1,
\gamma(0) = 0$ and $\Delta$, respectively, assumes positive
($\Delta > 0$), null ($\Delta = 0$) and negative ($\Delta < 0$) values:
\begin{center}
  \includegraphics[scale=0.45]{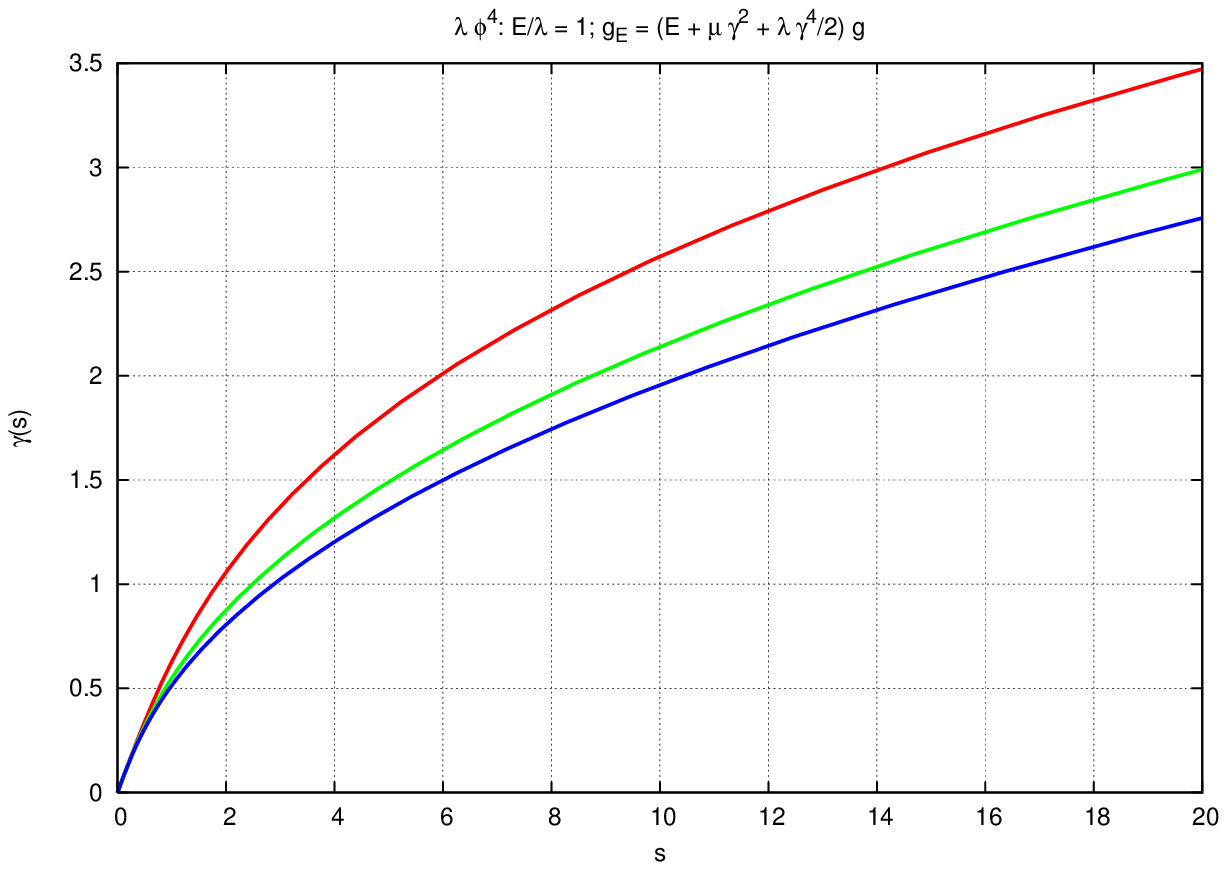} \hfill
  \includegraphics[scale=0.45]{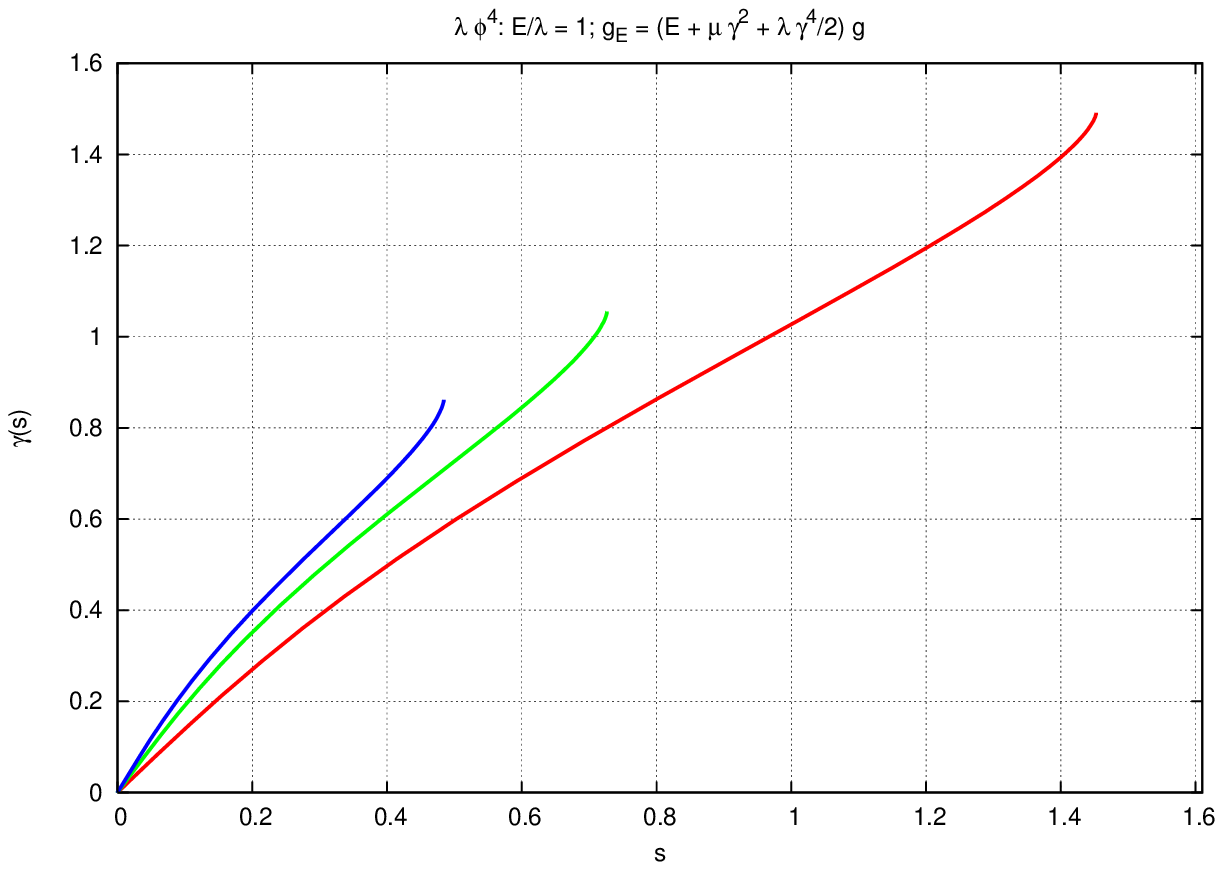}

  \includegraphics[scale=0.45]{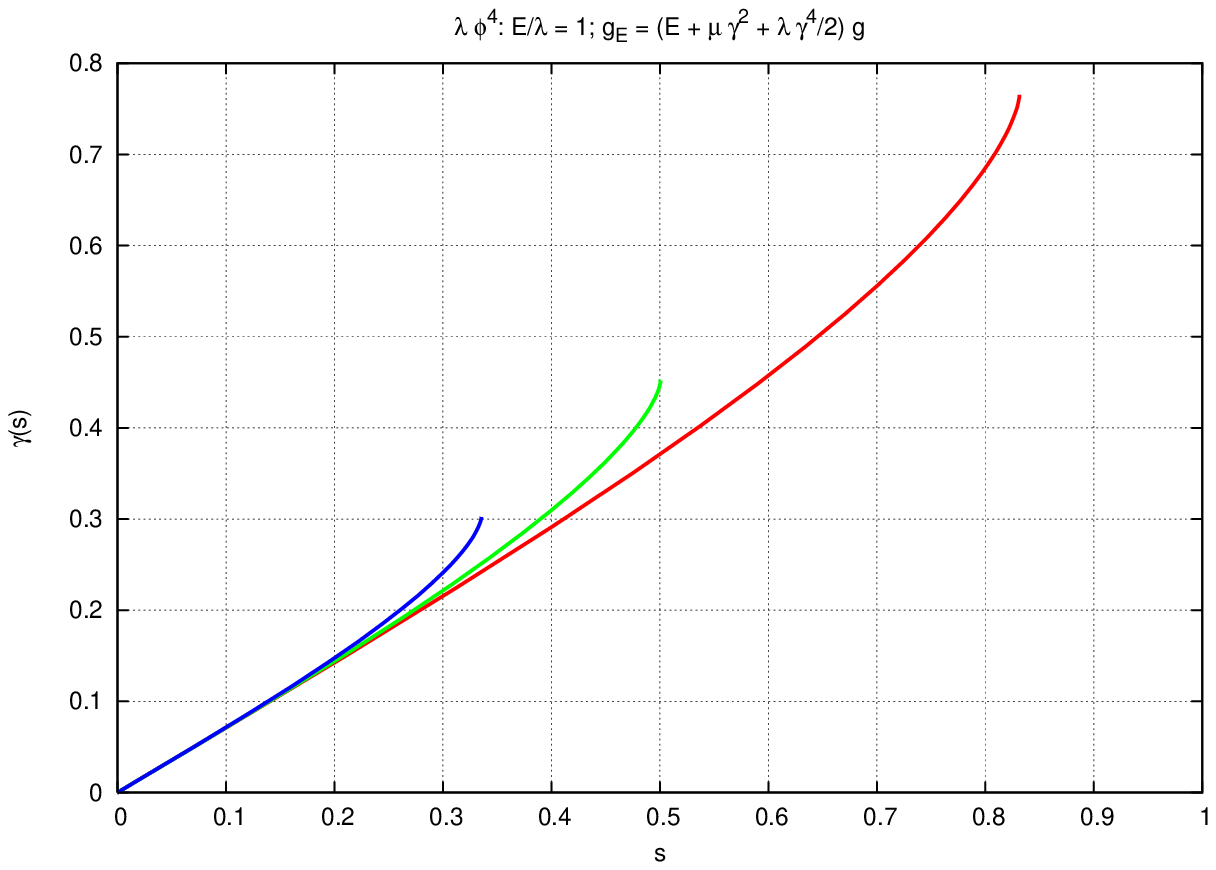}

  \bigskip \bigskip \bigskip
  {\footnotesize \noindent\textbf{Figure 3:} The top-left-corner plot shows the
    geodesic for $E/\lambda = 1$ and $\Delta > 0$, while the top-right-corner
    one depicts it for $E/\lambda = 1$ and $\Delta = 0$. The bottom-center plot
    shows $E/\lambda = 1$ and $\Delta < 0$.}
\end{center}
\bigskip \bigskip

We can clearly see that, for positive $\Delta$ (leftmost graph), we have a
smooth geodesic $\gamma(s)$ representing the symmetric phase. Then, when
$\Delta$ vanishes (middle graph), there is a clear change which delimits
the 2 different phases of the theory. Moreover, when $\Delta$ is negative
(rightmost graph), we have the third [broken-symmetric] phase of the theory
(which has a finite geodesic).
\section{Gauge Theory Examples}\label{sec:examples}
Now, we will consider examples of theories with gauge symmetry (rather than the
discrete $\mathbb{Z}_2$-symmetry of above). As seen in \eqref{eq:lp4geq},
taking $\gamma \mapsto -\gamma$ did not change the geodesic structure, i.e., the
procedure was covariant (resp. equivariant). This should come as no surprise,
once the potential, $V_{\tau}(\phi)$, is an equivariant function of the fields
and the spacetime (base manifold) metric does not depend on the gauge symmetry
involved either.

As mentioned before in section \ref{sec:appsqft}, we have assumed thus far that
the quantum fluctuations vanish at the boundary,
$\delta\phi|_{\partial\mathpzc{M}} = 0$. This, in turn, implies that the Jacobi
fields $\tilde{\pi}$ will satisfy $\tilde{\pi}|_{\partial\mathpzc{M}} =
0$. While the fields considered have no internal structure (i.e., gauge
symmetry), this is a fairly straightforward constraint, in the sense that its
solution is trivial (albeit labeled by $\tau$). However, when gauge symmetry is
present, there may be \emph{non-trivial} solutions to these constraints. This is
the phenomenon of spontaneous symmetry breaking.

Therefore, following the discussion done in section \ref{sec:appsqft}, the
following may happen: the quantum corrections, which are handles attached to the
classical solution, may now be able to change the topology of the classical
solution. So, when a symmetry is reduced from $G \rightarrow H$ (where $H$ is a
subgroup of $G$; see, e.g., \cite{ladies, kerbrat, fulpnorris, mayer,
  sardanashvily, neeman}), we start with a Jacobi field which satisfies
$\tilde{\pi}_{G}|_{\partial\mathpzc{M}} = 0$ and end up with a Jacobi field who only has
$H$ as symmetry, which implies that $\tilde{\pi}_{H}|_{\partial\mathpzc{M}} = 0$ ---
this means that the initial degrees-of-freedom that combined (respecting the $G$
symmetry) to yield the [initial] constraint $\tilde{\pi}_{G}|_{\partial\mathpzc{M}} =
0$, are not all available now, such that only part of the original symmetry is
still respected, yielding $\tilde{\pi}_{H}|_{\partial\mathpzc{M}} = 0$ (the remaining
degrees-of-freedom having recombined in non-trivial ways); thus, we end up with
a source (or sink) of Jacobi fields that only have $H$ as symmetry.

In this sense, the Atiyah-Singer Index Theorem can be used to measure this
variation (i.e., to measure the inequivalent representations of the algebra of
observables):

\begin{enumerate}
\item If all of the quantum corrections preserve the topology of the classical
  solution, the topological index does not change, which implies that the
  analytical index of the differential operator in question also does not
  change, which, in turn, leaves the vacuum state unchanged.
\item On the other hand, if the quantum fluctuations change the topology of the
  classical solution (as described above), the topological index will change
  (following the attachment of the appropriate $\nu$-cell), implying that the
  analytical index changes as well, which means that the vacuum state changes.
\end{enumerate}

From a different viewpoint, the question can be posed in the following way:
Given a certain symmetry breaking connection, how can the topology of the moduli
space of the associated Higgs Bundle be studied?

There are some studies in this direction, but no general answers: this is
because these types of characterizations are highly model-dependent, i.e., they
depend on the particular properties of the connection chosen for the Higgs
Bundle in question. For instance, following the discussion above, if the
boundary $\partial\mathpzc{M}$ is non-existent (i.e., $\mathpzc{M}$ is compact),
then $\tilde{\pi}|_{\partial\mathpzc{M}} = 0$ is trivialy satisfied; however, if
$\mathpzc{M}$ has, e.g., punctures (i.e., a finite set of points
omitted), then the connection in question will have a certain ramification
structure and the constraint $\tilde{\pi}|_{\partial\mathpzc{M}} = 0$ will have
non-trivial solutions (see, e.g., \cite{msheld}, and references therein);
therefore, the boundary conditions (in the sense of \cite{garciaguralnik,
  mollifier}) on the Jacobi fields ultimately determine the structure of the
connection (i.e., gauge field).

The examples below are twofold: the first one (Landau-Ginzburg Functional,
subsection \ref{subsec:lgf}) serves the purpose of showing the equivariance of
the method discussed in this work, once the connection (which is
$\mathfrak{u}(1)$-valued) does not undergo symmetry breaking; while the second
one (Seiberg-Witten Functional, subsection \ref{subsec:swf}) can be understood
as defined for pairs $(A, \phi)$, where $A$ is a Hermitian connection
(compatible with the holomorphic estructure on the bundle in question) and $\phi$
is a section (of the bundle at hand) --- in this sense, one can consider the
space of solutions to Hitchin's equations (i.e., the moduli space of the
associated Higgs Bundle) given by,

\begin{align}
  \label{eq:hitchineq1}
  F_A + [\phi, \phi^{*}] &= 0 \; ; \\
  \label{eq:hitchineq2}
  d_A'' \phi &= 0 \; ;
\end{align}
where $F_A$ is the curvature of $A$ and $d_A'' \phi$ is the anti-holomorphic
part of the covariant derivative of $\phi$ (compare these with
equations \eqref{eq:sw3} and \eqref{eq:sw4}). In this sense, the different
solutions found via \eqref{eq:swjacobi} are a direct statement about the
topology of the moduli space of the Higgs Bundle under study.
\subsection{The Landau-Ginzburg Functional}\label{subsec:lgf}
To start off, let us consider the case where the base manifold is a compact Riemann
surface $\Sigma$ equipped with a conformal metric and the vector bundle is a Hermitian
line bundle $\mathpzc{L}$ (i.e., with fiber $\mathbb{C}$ and a Hermitian metric
$\hm{\cdot}{\cdot}$ on the fibers).

The Landau-Ginzburg functional is defined for a section $\varphi$ and a unitary connection
$D_A = \ed\, + A$ of $\mathpzc{L}$ as ($\sigma\in\mathbb{R}$ is a real scalar),

\begin{equation*}
  L(\varphi,A) = \int_{\Sigma} |F_A|^2 + |D_A\, \varphi|^2 + \frac{1}{4}\, \bigl(\sigma -
    |\varphi|^2\bigr)^2 \; .
\end{equation*}

Thus, its Euler-Lagrange equations [of motion] are given by:
\begin{align*}
  D_A^{*}\, D_A\, \varphi &= \frac{1}{2}\, \bigl(\sigma - |\varphi|^2\bigr)\, \varphi\;;\\
  D_A^{*}\, F_A &=  -\re\hm{D_{A}\, \varphi}{\varphi} \; ;
\end{align*}
where, $D_A^{*}$ is the dual of $D_A$, i.e., $D_A^{*} = -*\, D_A\, * = -*\, (\ed\,
+ A)\, *$. Note that the equation for $F_A$ (the second one above) is linear 
in $A$. Since $D_A$ is a unitary connection, $A$ is a $\mathfrak{u}(1)$-valued
1-form. This Lie algebra (of the group $U(1)$) will sometimes be identified with
$i\, \mathbb{R}$ --- in other words, our $A$ corresponds to $-i\, A$ in the
standard physics literature (where $A$ is real-valued).

Before we go any further, some notational remarks are in order. We decompose the
space of 1-forms, $\Omega^1$, on $\Sigma$ as $\Omega^1 = \Omega^{1,0} \oplus
\Omega^{0,1}$, with $\Omega^{1,0}$ spanned by 1-forms of the type $dz$ and
$\Omega^{0,1}$ by 1-forms of the type $d\bar{z}$. Here $z = x + i\, y$ is a
local conformal parameter on $\Sigma$ and $\bar{z} = x - i\, y$. Therefore, $dz
= dx + i\, dy$, $d\bar{z} = dx - i\, dy$, $\partial_z = \tfrac{1}{2}\,
(\partial_x - i\, \partial_y)$ and $\partial_{\bar{z}} = \tfrac{1}{2}\,
(\partial_x + i\, \partial_y)$. Furthermore, if $\partial_x$ and $\partial_y$
are an orthonormal basis of the tangent space of $\Sigma$ at a given point, we
have that $\hm{dz}{dz} = 2$, $\hm{d\bar{z}}{d\bar{z}} = 2$ and
$\hm{dz}{d\bar{z}} = 0$. Given that the decomposition $\Omega^1 = \Omega^{1,0}
\oplus \Omega^{0,1}$ is orthogonal, we may also decompose $D_A$ accordingly:
$D_A = \partial_A + \bar{\partial}_A$, where $\partial_A\varphi \in
\Omega^{1,0}(\mathpzc{L})$ and $\bar{\partial}_A\varphi \in
\Omega^{0,1}(\mathpzc{L})$ for all sections $\varphi$ of $\mathpzc{L}$
(holomorphic, $\bar{\partial}_A\, f(z,\bar{z}) = 0 \Leftrightarrow
f(z,\bar{z}) = f(z)$, and anti-holomorphic, $\partial_A f(z,\bar{z}) = 0
\Leftrightarrow f(z,\bar{z}) = f(\bar{z})$, parts; i.e., the space of 1-forms
and the space of connections is decomposable into a direct sum of its
holomorphic and anti-holomorphic parts: $\partial_A = \partial + A^{1,0}$ and
$\bar{\partial}_A = \bar{\partial} + A^{0,1}$; while the exterior derivative is
given by $\ed\, = \partial + \bar{\partial}$). As expected, we have that
$\partial_A\, \partial_A = 0 = \bar{\partial}_A\, \bar{\partial}_A$ and $F_A =
-\bigl(\partial_A\, \bar{\partial}_A + \bar{\partial}_A\, \partial_A\bigr)$.

It is not difficult to show \cite{geomana} that:

\begin{align*}
    L(\varphi,A) &= \int_{\Sigma} |F_A|^2 + |D_A\, \varphi|^2 + \frac{1}{4}\, \bigl(\sigma -
      |\varphi|^2\bigr)^2 \; ; \\
    &= 2\, \pi\, \deg(\mathpzc{L}) + \int_{\Sigma} 2\, \bigl|\bar{\partial}_A\,
      \varphi\bigr|^2 + \Bigl(*(-i\, F_A) - \frac{1}{2}\, \bigl(\sigma -
      |\varphi|^2\bigr)\Bigr)^2 \; ; \\
    \text{where}\quad \deg(\mathpzc{L}) &= c_1(\mathpzc{L}) = \frac{i}{2\, \pi}\, \tr(F_A)\; ;
\end{align*}
i.e., the degree of the line bundle is given by the 1st Chern class.

Therefore, a very useful consequence from the above is that if $\deg(\mathpzc{L})
\geqslant 0$ the lowest possible value for $L(\varphi,A)$ is realized if $\varphi$ and $A$
satisfy

\begin{align}
  \label{eq:lg1}
  \bar{\partial}_A \, \varphi &= 0 \; ;\\
  \label{eq:lg2}
  *(i\, F_A) &= \frac{1}{2}\, \bigl(\sigma - |\varphi|^2\bigr) \; .
\end{align}

These equations are just the expression of the self-duality of the
Landau-Ginzburg functional. If $\deg(\mathpzc{L}) < 0$, then these equations
cannot have any solution, thus one has to consider the self-duality equations
arising from the Landau-Ginzburg functional where the term $+2\, \pi\,
\deg(\mathpzc{L})$ is substituted by $\boldsymbol{-}2\, \pi\,
\deg(\mathpzc{L})$. Therefore, without loss of generality, we shall assume
$\deg(\mathpzc{L}) \geqslant 0$. A necessary condition for the
solvability of \eqref{eq:lg2} is that,

\begin{align}
  \nonumber
  2\, \pi\, \deg(\mathpzc{L}) = \int i\, F_A &= \frac{1}{2}\, \int_{\Sigma} \bigl(\sigma -
    |\varphi|^2\bigr) \leqslant \frac{\sigma}{2}\, \text{Area}(\Sigma)\; ; \\
  \label{eq:lgentropy}
  \therefore\; \sigma &\geqslant \frac{4\, \pi \deg(\mathpzc{L})}{\text{Area}(\Sigma)} \; ;
\end{align}
and the equality only occurs if, and only if, $\varphi \equiv 0$.

The following result is useful when studying the solutions of the above
functionals \cite{geomana}: \emph{Let $\Sigma$ be a compact Riemann surface with
  a conformal   metric and $\mathpzc{L}$ as before. For any solution of
  \eqref{eq:lg1}, we have that $|\varphi| \leqslant \sigma$ on $\Sigma$.} That
is, this maximum principle states that the amplitude of the field cannot exceed
the height of the potential (see the first plot below).

Let us now construct the Jacobi metric for this potential and find its possible
geodesics:
\begin{align*}
  \gE &= 2\, \big(E - V_{\sigma}(\gamma)\big)\, \g \; ; \\
  &= 2\, \bigg(E + \frac{1}{4}\,\big(\sigma - |\gamma|^2\big)^2\bigg)\, \g \; .
\end{align*}

In a complete analogy to what was previously done, we consider the $P(\gamma) =
E - V(\gamma)$ polynomial. However, in order to make this analysis clearer, let
us, in fact, use a slightly different polynomial given by $P^{\prime}(\gamma) =
4\, \bigl(P(\gamma) - E\bigr)$. It is straightforward to see that $P'(\gamma) =
\bigl(|\gamma|^2 - \sigma\bigr)\, \bigl(|\gamma|^2 - \sigma\bigr)$, which means
that $\sigma$ is the only root of $P'(\gamma)$, with double multiplicity (the
two roots coalesce into one).

This situation implies that the discriminant of $P'(\gamma)$ vanishes, i.e.,
$\Delta = 0$, and the different phases of the theory are labelled by $\sigma =
0$ and $\sigma > 0$.

The plots below have, respectively, the following
values for $\sigma$: $0.0$, $3.0$, $5.0$, $7.0$, $11.0$ and $31.0$. As they
show, when $\sigma = 0$ we have $c_1(\mathpzc{L}) = \deg(\mathpzc{L}) =
\frac{i}{2\, \pi}\, \tr(F_A) = 0$ and its geodesic has a clear character which
is quite different otherwise:

\begin{center}
  \includegraphics[scale=0.4]{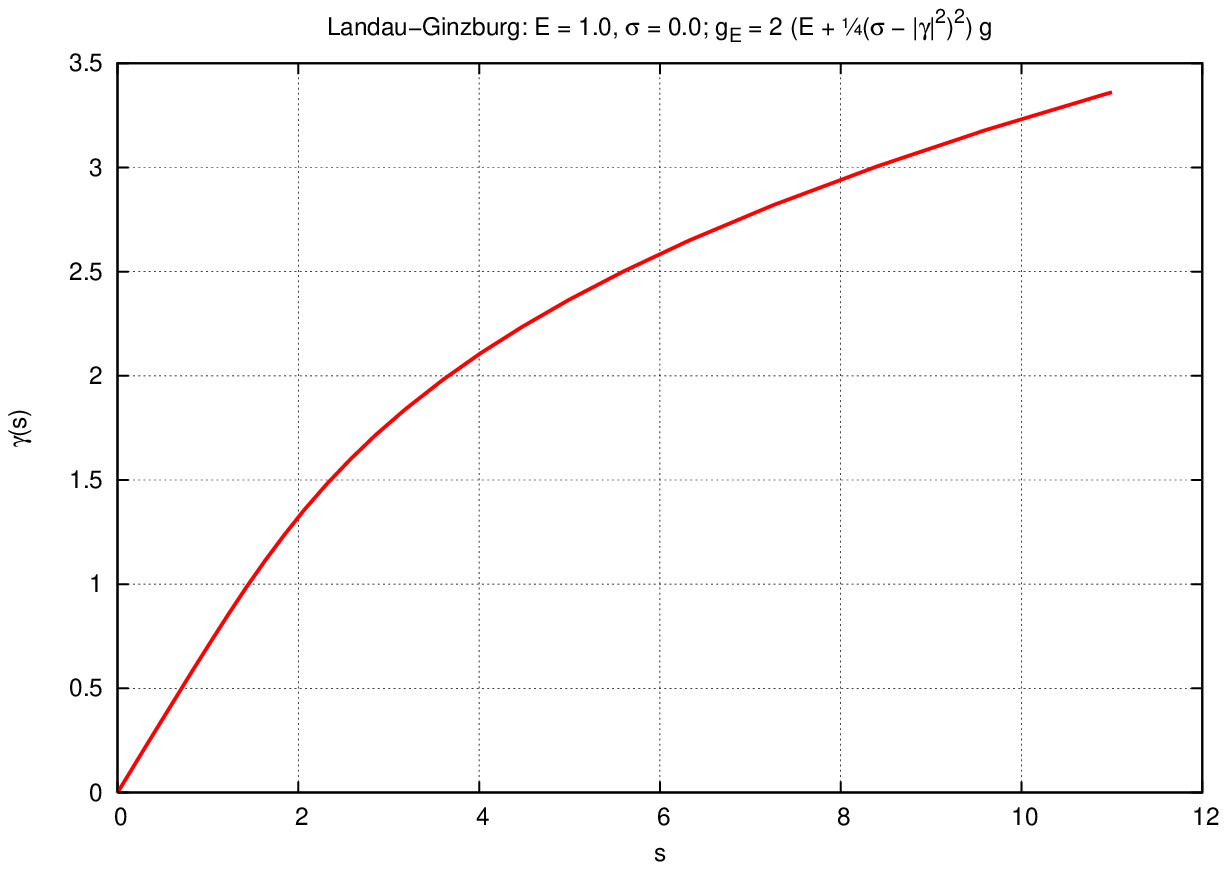} \hfill
  \includegraphics[scale=0.4]{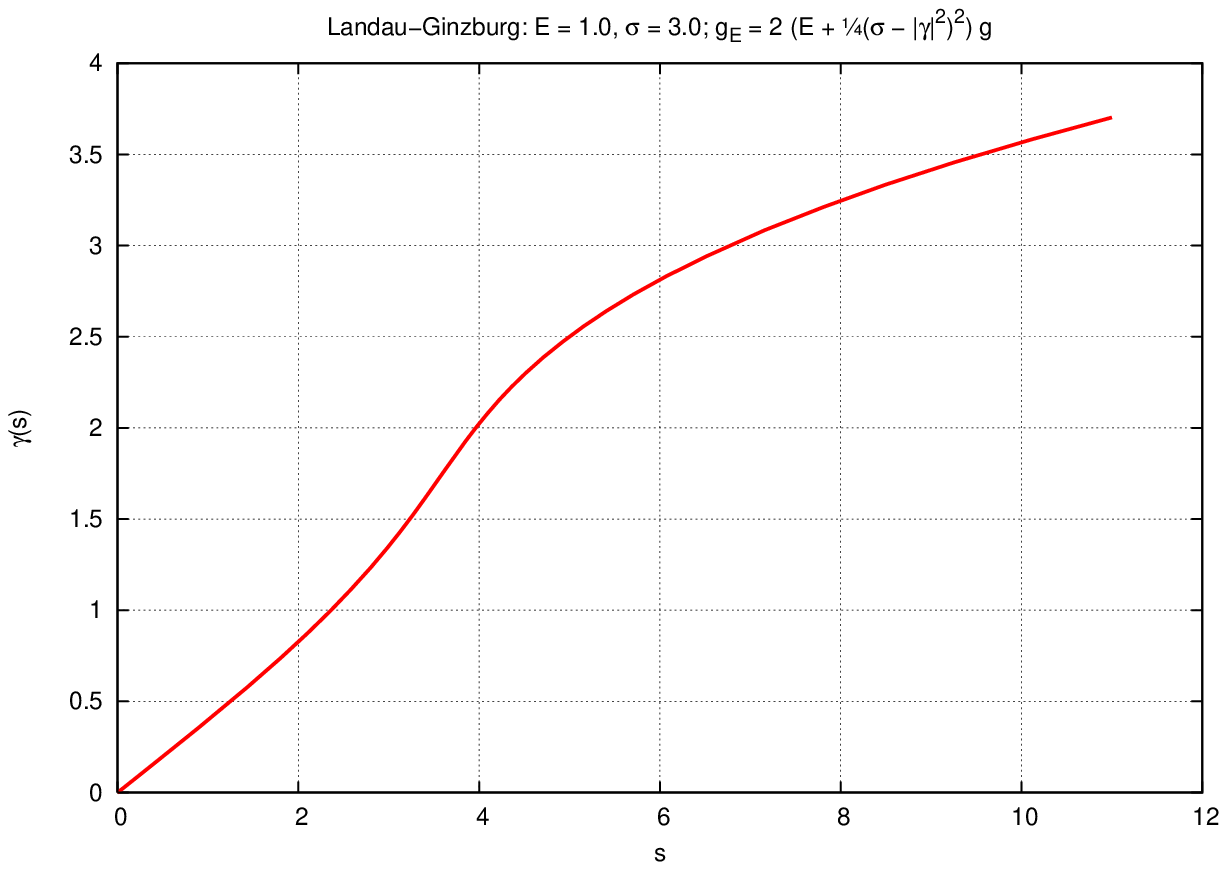} \hfill
  \includegraphics[scale=0.4]{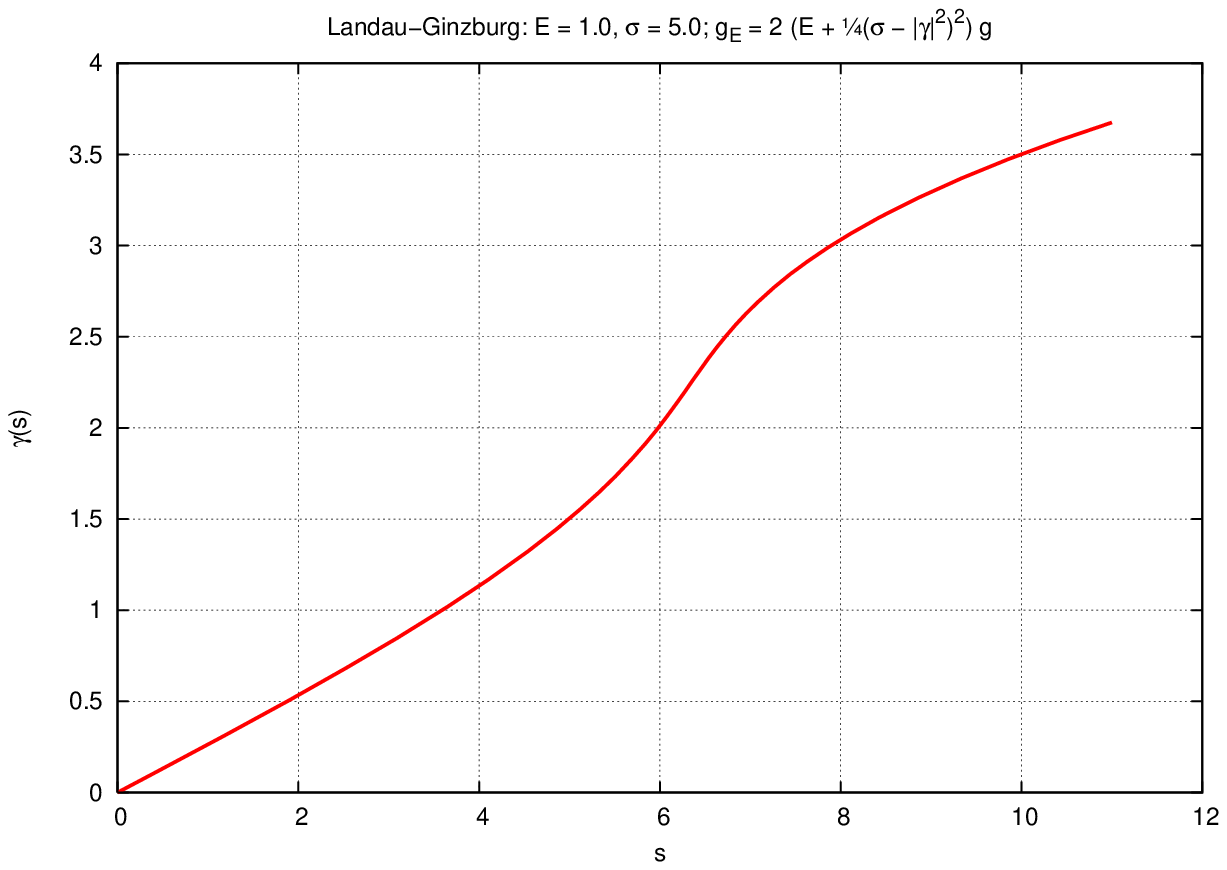} \hfill
  \includegraphics[scale=0.4]{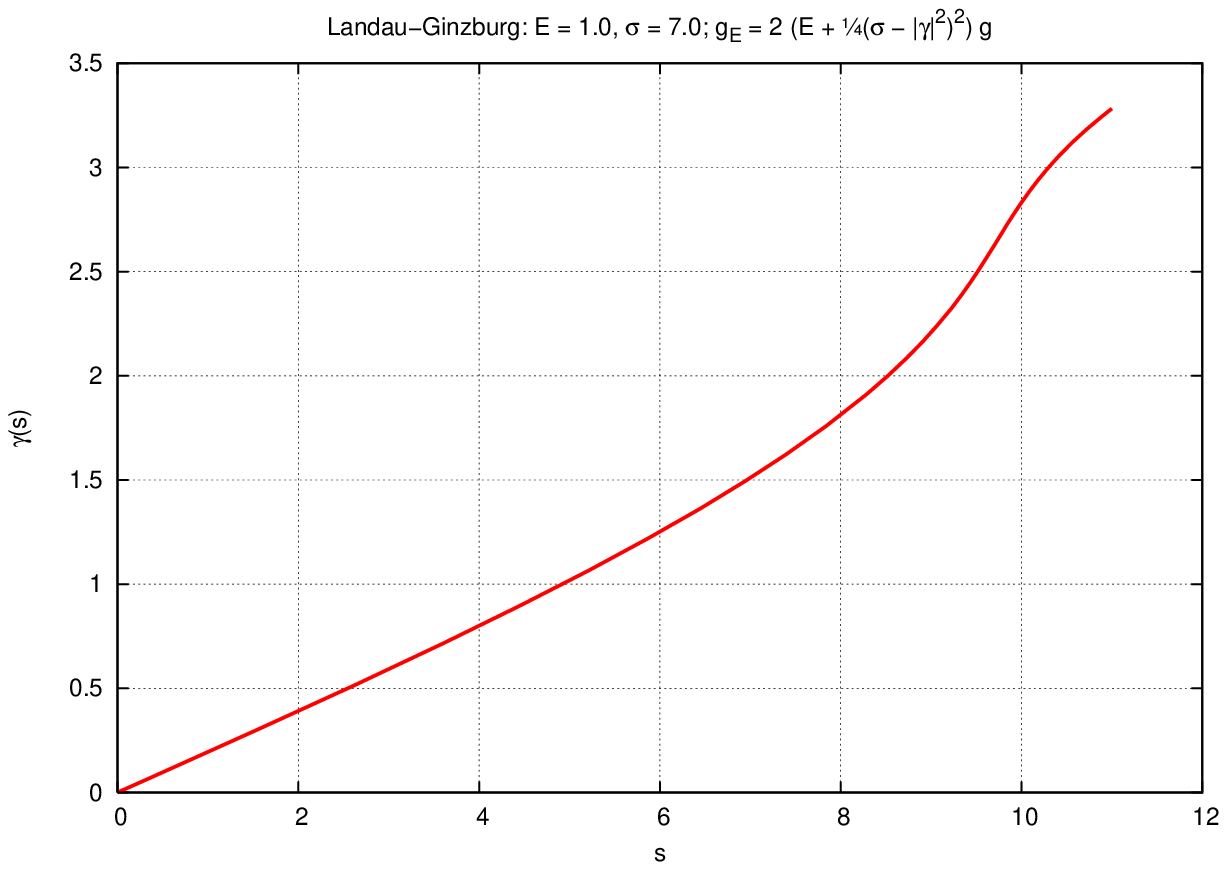} \hfill
  \includegraphics[scale=0.4]{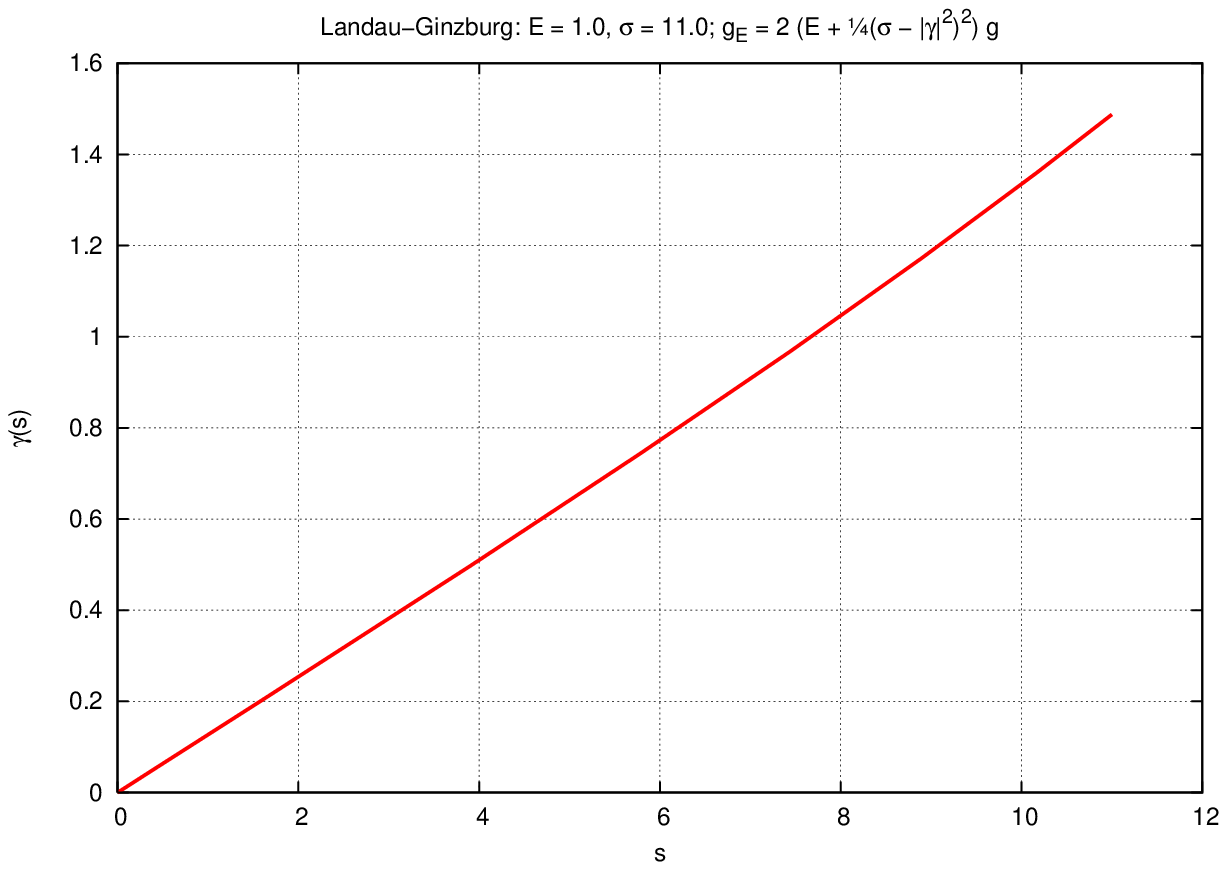} \hfill
  \includegraphics[scale=0.4]{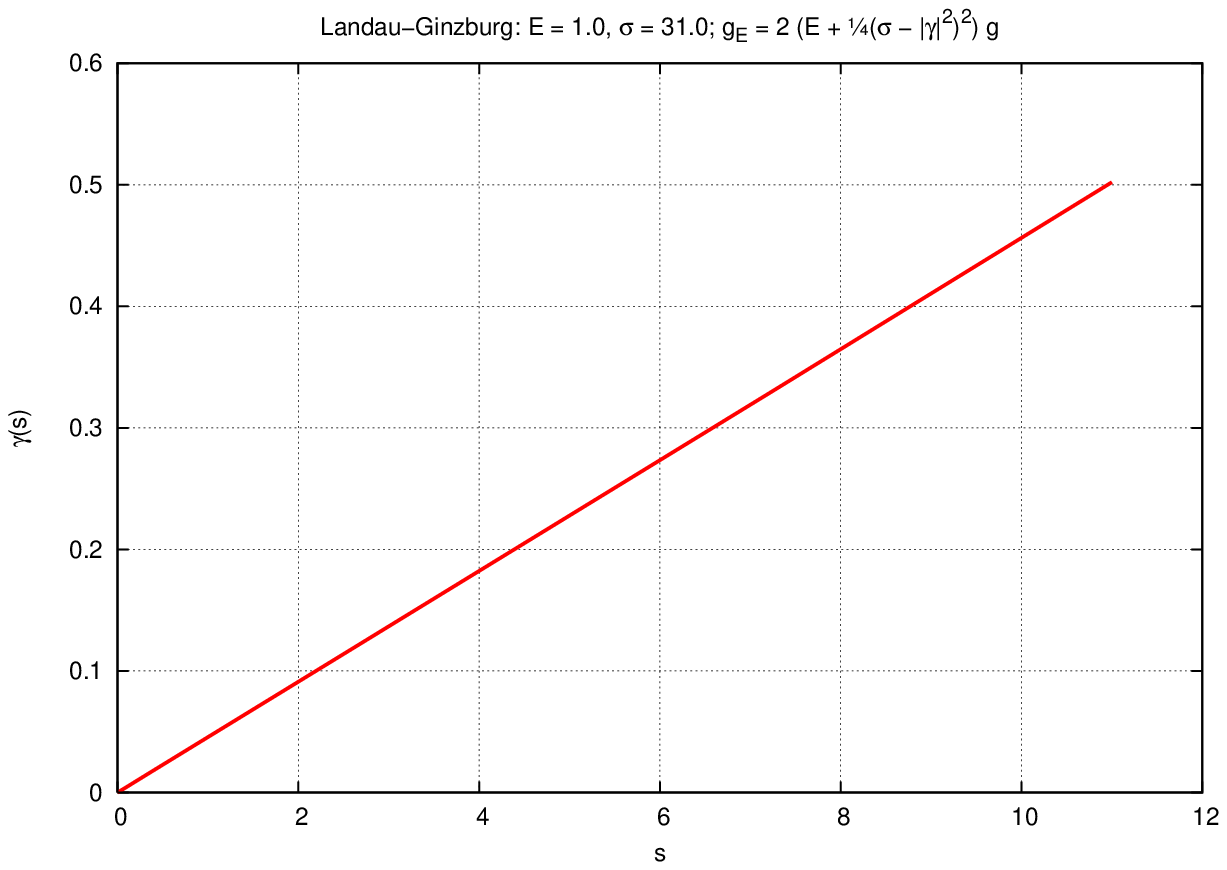} \hfill

  \bigskip \bigskip \bigskip
  {\footnotesize \noindent\textbf{Figure 4:} The plots above show the clear
    distinction between the $\sigma = 0$ solution and the $\sigma > 0$ ones; the
    first plot, on the upper left corner, has $\sigma = 0$, while the other ones
    have differing positive values for $\sigma$.}
\end{center}

As a last remark on this model, note it can be thought as defined by a
pair $(A, \varphi)$, where $A$ is a unitary connection compatible with the
holomorphic structure of the bundle in question, and $\varphi$ is a [global]
section of this bundle. In this sense, Hitchin's self-duality equations become,
\begin{align}
  \label{eq:hitchinlg1}
  \bar{\partial}_A \varphi &= 0 \; ; \\
  \label{eq:hitchinlg2}
  F_A + [\varphi, \varphi^{*}] &= 0 \; .
\end{align}
Compare these with \eqref{eq:lg1} and \eqref{eq:lg2}.

In this sense, the solutions found above are a direct statement about the
topology of the Higgs Bundle in question.
\subsection{The Seiberg-Witten Functional}\label{subsec:swf}
In this case, the base manifold, $\mathpzc{M}$, is a compact, oriented,
4-dimensional Riemannian manifold endowed with a spin${}^c$ structure, \ie, a
spin${}^c$ manifold. The determinant line of this spin${}^c$ structure will be
denoted by $\mathpzc{L}$ and the Dirac operator determined by a unitary
connection $A$ on $\mathpzc{L}$ will be denoted by $\mathcal{D}_A$. Recalling
the half spin bundle $\mathcal{S}^{\pm}$ defined by the spin${}^c$ structure, we
see that $\mathcal{D}_A$ maps sections of $\mathcal{S}^{\pm}$ into sections of
$\mathcal{S}^{\mp}$ \cite{seibergwitten,donaldson,liviu,geomana}.

In this fashion the Seiberg-Witten functional for a unitary connection $A$ on
$\mathpzc{L}$ and a section $\varphi$ of $\mathcal{S}^+$ is given by,

\begin{equation}
  \label{eq:swf}
  SW[\varphi,A] = \int_{\mathpzc{M}} |\nabla_A \varphi|^2 + |F^+_A|^2 + \frac{R}{4}\,
    |\varphi|^2 + \frac{1}{8}\, |\varphi|^4 \; ;
\end{equation}
where $\nabla_A$ is the spin${}^c$ connection induced by $A$ and the Levi-Civita
connection of $\mathpzc{M}$, $F_A^+$ is the self-dual part of the curvature of $A$ and $R$
is the scalar curvature of $\mathpzc{M}$. Its Euler-Lagrange equations are given by,

\begin{align}
  \label{eq:sw1}
  \nabla_A^*\, \nabla_A\, \varphi &= -\biggl(\frac{R}{4} + \frac{1}{4}\, |\varphi|^2
    \biggr)\, \varphi \; ; \\
  \label{eq:sw2}
  \ed{}^*\, F_A^+ &= -\re\hm{\nabla_{A}\, \varphi}{\varphi} \; .
\end{align}

Using a spin frame, it is not difficult \cite{geomana} to show that the Seiberg-Witten
functional can be written in the following form:

\begin{equation*}
  SW[\varphi,A] = \int_{\mathpzc{M}} |\mathcal{D}_A \varphi|^2 + \biggl|F_A^+ -
    \frac{1}{4}\, \hm{e_j\cdot e_k\cdot \varphi}{\varphi}\, e^j \wedge e^k\biggr|^2 \; ;
\end{equation*}
where $e^j$ are 1-forms dual to the tangent vectors $e_j$, $e^j(e_k) = \delta^j_k$;
$j,k=1,\dotsc,4$. As a corollary of the above, the lowest possible value of the
Seiberg-Witten functional is achieved if $\varphi$ and $A$ are solutions of the
\emph{Seiberg-Witten equations}:

\begin{align}
  \label{eq:sw3}
  \mathcal{D}_A \varphi &= 0 \; ;\\
  \label{eq:sw4}
  F_A^+ &= \frac{1}{4}\, \hm{e_j\cdot e_k\cdot \varphi}{\varphi}\, e^j \wedge e^k \; .
\end{align}

Thus, self-duality is at work yet again: the absolute minima of the Seiberg-Witten
functional satisfy not only the second order equations \eqref{eq:sw1} and \eqref{eq:sw2},
but also the first order Seiberg-Witten equations \eqref{eq:sw3} and \eqref{eq:sw4}.

Although our discussion of the Seiberg-Witten functional, so far, has mirrored our
discussion of the Landau-Ginzburg one, the parameter $\sigma$ on the latter has had no
analogue in the former. This can be accomplished with the introduction of a 2-form $\mu$
and the consideration of the perturbed functional,

\begin{align*}
  SW_{\mu}[\varphi,A] &= \int_{\mathpzc{M}} |\mathcal{D}_A \varphi|^2 + \biggl|F_A^+ -
    \frac{1}{4}\, \hm{e_j\cdot e_k\cdot \varphi}{\varphi}\, e^j \wedge e^k +
    \mu\biggr|^2 \; ; \\
  &= \int_{\mathpzc{M}} |\nabla_A \varphi|^2 + |F^+_A|^2 + \frac{R}{4}\,
    |\varphi|^2 + \biggl|\mu - \frac{1}{4}\, \hm{e_j\cdot e_k\cdot \varphi}{\varphi}\, e^j
    \wedge e^k\biggr|^2 + 2\, \hm{F_A^+}{\mu} \; .
\end{align*}
Their corresponding first order equations of motion are,

\begin{align*}
  \mathcal{D}_A \varphi &= 0 \; \\
  F^+_A &= \frac{1}{4}\, \hm{e_j\cdot e_k\cdot \varphi}{\varphi}\, e^j\wedge e^k - \mu\;.
\end{align*}

If we assume that $\mu$ is closed and self-dual, then we see that $\hm{F_A}{\mu} =
\hm{F_A^+}{\mu}$, once $\hm{F_A^-}{\mu} = 0$ due to the orthogonality between
anti-self-dual and self-dual forms. Thus, given that $F_A$ represents the first Chern
class $c_1(\mathpzc{L})$ of the line bundle $\mathpzc{L}$, and we assumed $\mu$ to be
closed (hence it represents a cohomology class $[\mu]$), the integral

\begin{equation*}
  \int_{\mathpzc{M}} \hm{F_A}{\mu} \; ,
\end{equation*}
does not depend on the connection $A$, thus representing a topological invariant, denoted
by $\bigl(c_1(\mathpzc{L})\wedge[\mu]\bigr)[\mathpzc{M}]$.

Just as before, we also have a maximum principle: \emph{For any solution $\varphi$ of
  \eqref{eq:sw1} --- in particular, for any solution of \eqref{eq:sw3} --- on a compact
  4-dimensional Riemannian manifold, we have that,}

\begin{equation*}
  \max_{\mathpzc{M}} |\varphi|^2 \leqslant \max_{x\in \mathpzc{M}} (-R(x), 0)\; .
\end{equation*}

As a direct consequence of this, if the compact, oriented, Riemannian Spin${}^c$ manifold
$\mathpzc{M}$ has nonnegative scalar curvature, the only possible solution of the
Seiberg-Witten equations is,

\begin{equation*}
  \varphi\equiv 0\; ; \quad F_A^+ \equiv 0 \; .
\end{equation*}

The Jacobi metric for this Seiberg-Witten model is given by

\begin{equation}
  \label{eq:swjacobi}
  \gE = 2\, \biggl(E + \frac{R}{4}\, |\gamma|^2 + \frac{1}{8}\, |\gamma|^4
    \biggr)\, \g \; ;
\end{equation}
for a geodesic $\gamma$; and, just like before, although $P(\gamma) = E -
V(\gamma) = E + R\, |\gamma|^2/4 + |\gamma|^4/8$, let us consider $P'(\gamma) =
8\, \bigl(P(\gamma) - E\bigr) = \bigl(|\gamma|^2 - 0\bigr)\, \bigl(|\gamma|^2 +
2\, R\bigr)$. It is then clear that when $R = 0$ the discriminant vanishes
$(\Delta = 0)$ and the 2 roots merge into 1, what constitutes one of the phases
of the theory. When $R \neq 0$, the discriminant is either $\Delta > 0$ $(R >
0)$ or $\Delta < 0$ $(R < 0)$, which accounts for the other 2 phases of the
theory.

The plots below were obtained with the choice of $E = 1.0$ and $R$ respectively
equal to $0.0$, $3.0$, $7.0$, $-3.0$, $-5.0$, $-7.0$.

\begin{center}
  \includegraphics[scale=0.4]{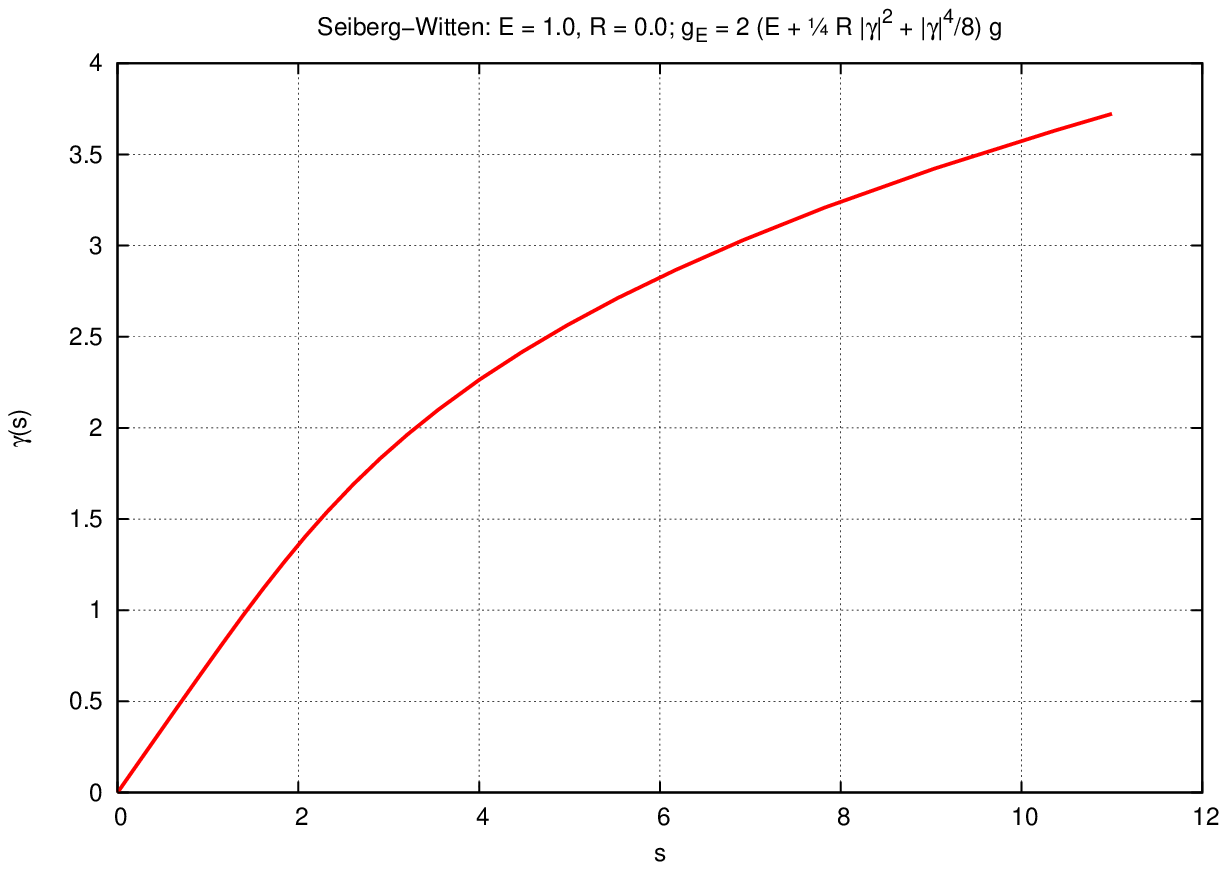} \hfill
  \includegraphics[scale=0.4]{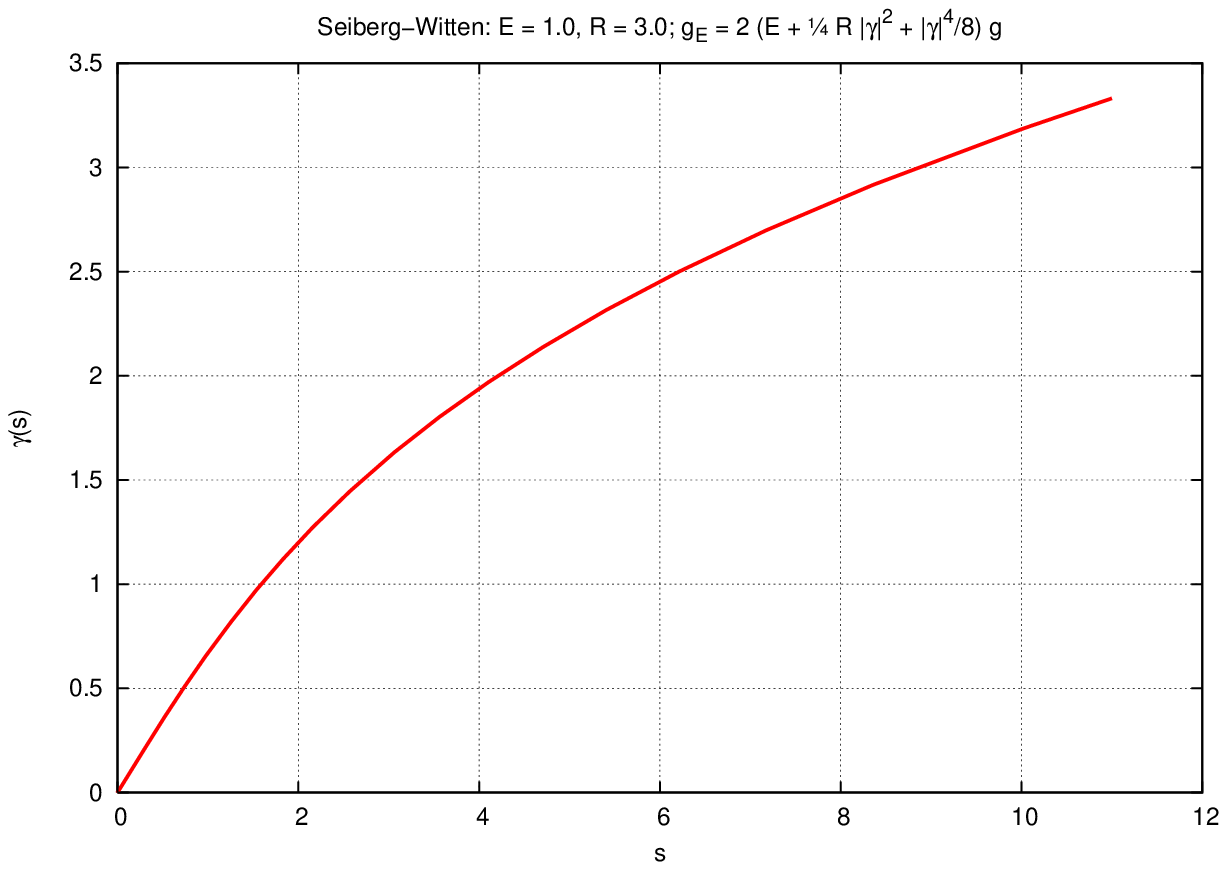} \hfill
  \includegraphics[scale=0.4]{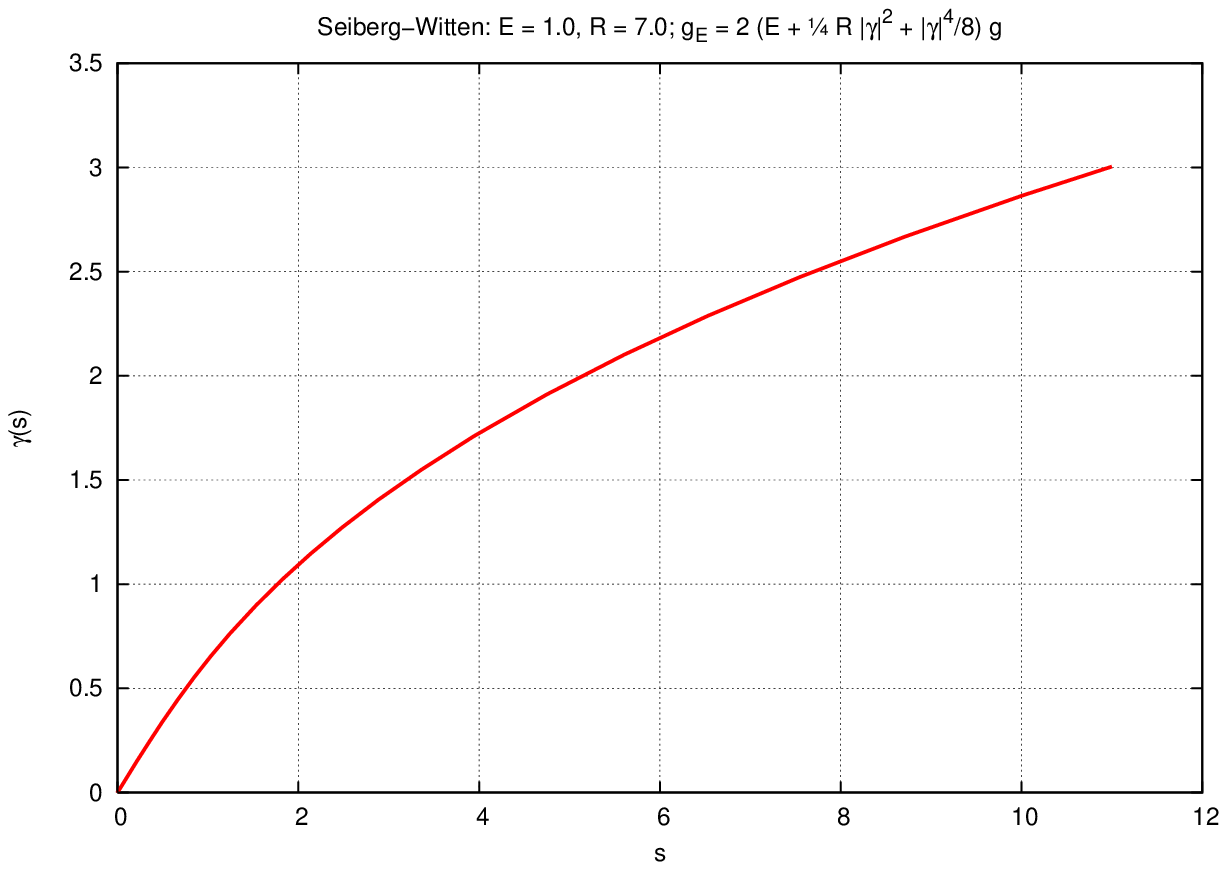} \hfill
  \includegraphics[scale=0.4]{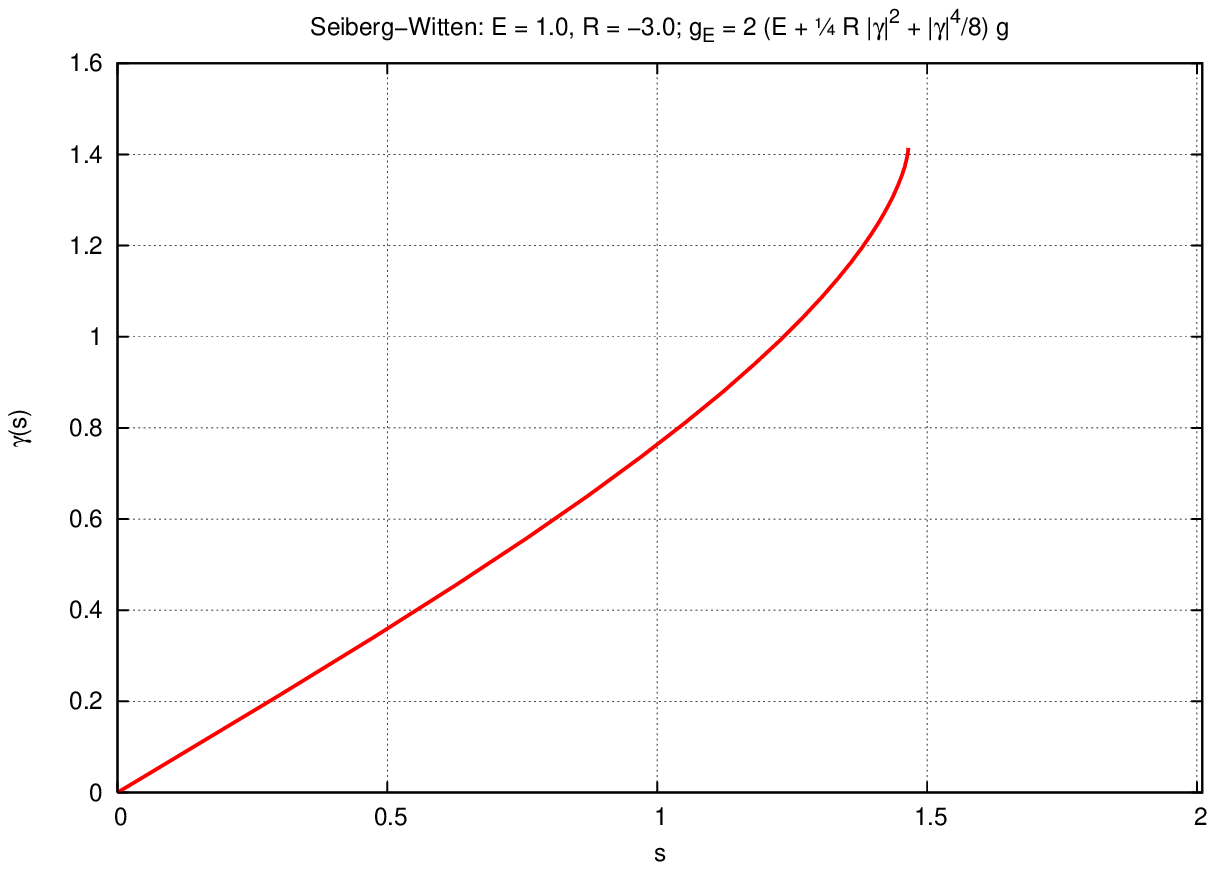} \hfill
  \includegraphics[scale=0.4]{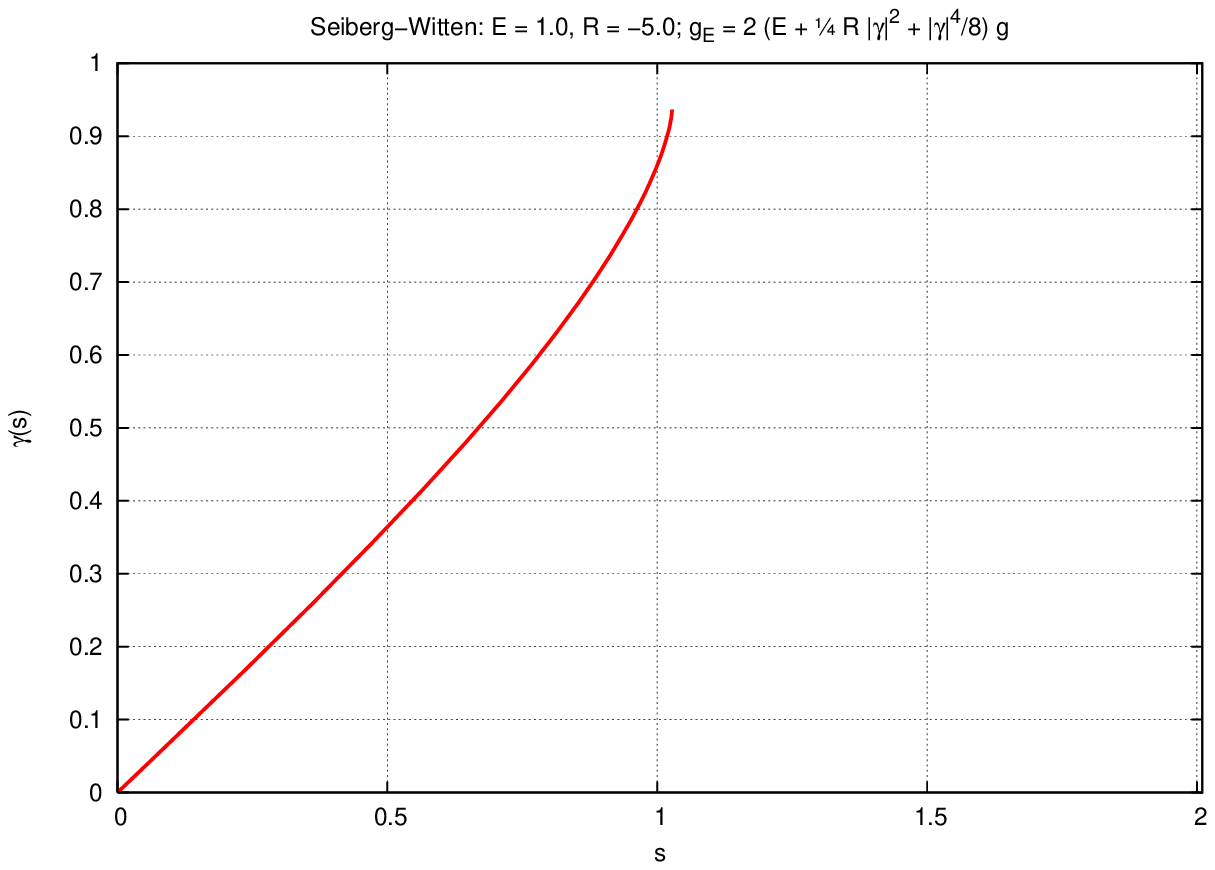} \hfill
  \includegraphics[scale=0.4]{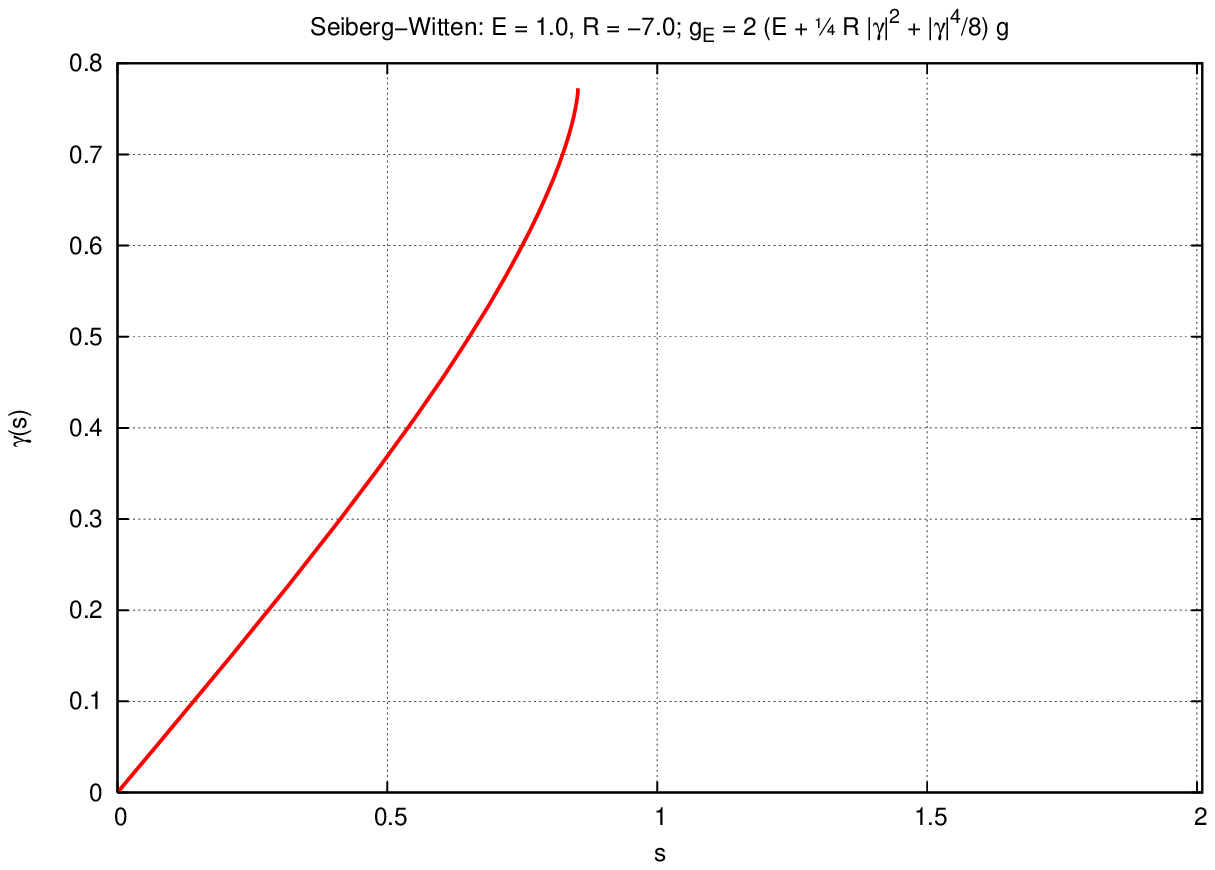} \hfill

  \bigskip \bigskip \bigskip
  {\footnotesize \noindent\textbf{Figure 5:} The plot in the upper left corner
    depicts the solution for $R=0$, while the plots in the upper center and
    upper right corner show solutions for $R > 0$. The plots in the lower row
    picture the solutions for $R < 0$.}
\end{center}
\bigskip
\section{Conclusions} \label{sec:conclusion}
Using a conformal transformation that amounts to finding the arc-length
reparameterization of the given problem, we were able to construct a new metric,
called Jacobi metric, such that its geodesic equation is equivalent to the
original equations of motion.

Then, we separated the given problem into its classical part and quantum
fluctuations and, by realizing that the action written in terms of the Jacobi
metric is the so-called ``energy'' in Morse Theory (resp. Morse-Bott theory), we
see can reinterpret this as a ``topological expansion'', in the sense that the
quantum fluctuations are handles attached to the classical solution.

Further, when these handles change the original (classical) topology, via the
gluing of an appropriate $\nu$-cell, there is a phase transition, in the sense
that we move from one solution [of our QFT] to another. This can be measured via
the use of an Index Theorem: once it relates the analytical index of the
differential operator in question with the topological index of the manifold
under study, we clearly see that if the topological index changes (because of a
certain handle attachment), the analytical index must change as well; which
means that the zero modes, the solutions of the equations of motion, have to
change.

Therefore, by studying the geodesics obtained from the Jacobi metric with
respect to its parameters, we can compute the topological index (via
generalizations of the Gauss-Bonnet theorem, using the Jacobi metric we derived
before) and thus compute the zero-mode solutions with respect to the different
values of the parameters. In this way, we are able to classify all possible
solutions to the QFT in question, keeping in mind that the quantum corrections
are given by handle attachments that may or may not (depending on the values of
the parameters) change the topology of the classical solutions.

To this picture, we add Lee-Yang zeros and Stokes phenomena, in order to obtain
a more robust view of what is at stake: the Lee-Yang zeros accumulate along
Stokes lines and pinch the parameter space, creating different regions of
``allowed values'' for the parameters of the theory (mass, coupling constants,
etc). This represents different sectors of theory, different phases of the
theory. And we showed that these are topologically inequivalent.

In this fashion, the partition function is meromorphic, once it is
singular along the Stokes' lines. However, it has various degress of modular
symmetry, depending on the particulars of the theory in question. And, in this
sense, different sectors of the theory are related to each other [by an
appropriate modular transformation].

In turn, this implies that different solutions of the theory are related to each
other, a fact that we dubbed ``duality''. Ultimately, these dualities are
determined by the actual values of the parameters (and how they compose in order
to create the modular symmetry in question), which are determined by the
boundary conditions of the Schwinger-Dyson equations (as explained earlier).

Future work will focus on D-modules and dimensional construction of
0-dimensional theories (which can be completely solved). We also intend to
generalize the cubic potential showed earlier to matrix- and Lie-algebra-valued
fields \cite{lieairy}, analytically solving the its 0-dimensional counterpart
and dimensionally constructing it (via D-modules): it seems plausible that such
an extension might be related to current developments in three-dimensional gravity,
\cite{3dgravity}.
\acknowledgments
The authors would like to thank C. Pehlevan for discussions and useful
conversations. This work is supported in part by funds provided by the US
Department of Energy (\textsf{DOE}) under contracts \textsf{DE-FG02-85ER40237}
and \textsf{DE-FG02-91ER40688-TaskD}.


\begin{thebibliography}{99}
  \bibitem{frankel} Theodore Frankel, \emph{The Geometry of Physics: An
      Introduction}, Cambridge University Press, 2003.
  \bibitem{nakahara} M. Nakahara, \emph{Geometry, Topology and Physics}, Taylor
    \& Francis, 2003.
  \bibitem{darling} R. W. R. Darling, \emph{Differential Forms and Connections},
    Cambridge University Press, 1994.
  \bibitem{ladies}  Y. Choquet-Bruhat, C. DeWitt-Morette, \emph{Analysis,
      Manifolds and Physics, Part I and II}, North
    Holland.
  \bibitem{kerbrat} Yvan Kerbrat, Helene Kerbrat-Lunc, \emph{Spontaneous
      symmetry breaking and principal fibre bundles}, \href{http://www.sciencedirect.com/science?_ob=ArticleURL&_udi=B6TJ8-46G4W9H-2Y&_coverDate=12%2F31%2F1986&_alid=3842
13658&_rdoc=1&_fmt=&_orig=search&_qd=1&_cdi=5304&_sort=d&view=c&_acct=C000022678&_version=1&_urlVersion=0&_userid=489286&md5=bff69edc9c9e393233dfbf1ff9be5172}{J. Geometry
 and Physics
    Vol. III, No. 2 (1986)}.
  \bibitem{fulpnorris} R. O. Fulp, L. K. Norris, \emph{Splitting of the
      connection in gauge theories with broken symmetry},
    \href{http://scitation.aip.org/getabs/servlet/GetabsServlet?prog=normal&id=JMAPAQ000024000007001871000001&idtype=cvips&gifs=yes}{J. of Mathematical Physics, Vol. 24, 
1871, (1983)}.
  \bibitem{mayer} M. E. Mayer, \emph{The geometry of symmetry breaking in gauge
      theories}, Act. Phys. Austr. (Suppl.) XXIII, 477--490 (1941).
  \bibitem{sardanashvily} Gennadi Sardanashvily, \emph{On the geometry of
      spontaneous symmetry breaking},
    \href{http://scitation.aip.org/getabs/servlet/GetabsServlet?prog=normal&id=JMAPAQ000033000004001546000001&idtype=cvips&gifs=yes}{J. Math. Phys. 33 (4), April 1992}.
  \bibitem{simpson} Carlos T. Simpson, \emph{Higgs bundles and local systems},
    \href{http://www.numdam.org/item?id=PMIHES_1992__75__5_0}{Publications
      Math{\'e}matiques de l'IH{\'E}S, 75 (1992), p. 5--95}.
  \bibitem{neeman} Y. Ne'eman, \emph{Geometrization of Spontaneously Broken
      Gauge Symmetries},
    \href{http://www.springerlink.com/content/p21828x959210654/}{Theoretical and
      Mathematical Physics, \textbf{139}(3): 745--750 (2004)}.
  \bibitem{symb} G. S. Guralnik, C. R. Hagen, T. W. B. Kibble, \emph{Global
      Conservation Laws and Massless Particles},
    \href{http://prola.aps.org/abstract/PRL/v13/i20/p585_1}{Phys. Rev. Lett. 13,
      585--587 (1964)}. P. W. Higgs, \emph{Broken symmetries, massless particles
      and gauge fields},
    \href{http://www.sciencedirect.com/science?_ob=ArticleURL&_udi=B6X44-46WP33M-12&_coverDate=09%2F15%2F1964&_alid=384209160&_rdoc=1&_fmt=&_orig=search&_qd=1&_cdi=7316&_
sort=d&view=c&_acct=C000022678&_version=1&_urlVersion=0&_userid=489286&md5=39e9e61ceeec1d93865305f227518bae}{Phys. Lett. \textbf{12} 132}. F. Englert and R. Brout, \emph{
Broken Symmetry and the Mass of Gauge Vector Mesons},
\href{http://prola.aps.org/abstract/PRL/v13/i9/p321_1}{Phys. Rev. Lett. 13,
  321--323 (1964)}. P. W. Higgs, \emph{Broken Symmetries and the Masses of Gauge
  Bosons},
\href{http://prola.aps.org/abstract/PRL/v13/i16/p508_1}{Phys. Rev. Lett. 13,
  508--509 (1964)}.
  \bibitem{strocchi} F. Strocchi, \emph{Spontaneous Symmetry Breaking in Local
      Gauge Quantum Field Theory; The Higgs Mechanism},
    \href{http://www.springerlink.com/content/u26gv53813601720/}{Commun. Math. Phys. 56, 57--78 (1977)}.
  \bibitem{weinbergcoleman} Sidney Coleman, Erick Weinberg, \emph{Radiative
      Corrections as the Origin of Spontaneous Symmetry Breaking},
    \href{http://prola.aps.org/abstract/PRD/v7/i6/p1888_1}{Phys. Rev. D 7,
      1888--1910 (1973)}. Erick Weinberg, \emph{Radiative Corrections as the
      Origin of Spontaneous Symmetry Breaking},
    \href{http://arxiv.org/abs/hep-th/0507214}{\texttt{hep-th/0507214}}.
  \bibitem{garciaguralnik} S. Garcia, G. Guralnik, Z. Guralnik, \emph{Theta
      Vacua and Boundary Conditions of the Schwinger Dyson Equations},
    \href{http://arxiv.org/abs/hep-th/9612079}{\texttt{hep-th/9612079}}. G. S. Guralnik,
    Z. Guralnik, \emph{Complexified Path Integrals and the Phases of Quantum
      Field Theory}, [\arXivid{0710.1256}].
  \bibitem{borcherds} R. E. Borcherds and A. Barnard, \emph{Lectures on Quantum
      Field Theory},
    \href{http://arxiv.org/abs/math-ph/0204014}{\texttt{math-ph/0204014}}.
  \bibitem{berrystokes} M. V. Berry, \emph{Stokes' phenomenon; smoothing a
      victorian discontinuity},
    \href{http://www.numdam.org/item?id=PMIHES_1988__68__211_0}{Publications
      Math{\'e}matiques de l'IH{\'E}S, 68 (1988),
      p. 211-221}. M. V. Berry, \emph{Uniform Asymptotic Smoothing of Stokes's
      Discontinuities},
    \href{http://www.jstor.org/stable/2398522}{Proc. R. Soc. Lond. A,
        Vol. 422, No. 1862 (Mar. 8, 1989), pp. 7-21}.
  \bibitem{mfmf} J. Milne, \emph{Modular Functions and Modular Forms},
    \href{http://www.jmilne.org/math/CourseNotes/math678.html}{\texttt{http://www.jmilne.org/math/CourseNotes/math678.html}}
  \bibitem{3dgravity} E. Witten, \emph{Three-Dimensional Gravity Revisited},
    [\href{http://arxiv.org/abs/0706.3359}{\texttt{arXiv:0706.3359}}]. A. Maloney,
    E. Witten, \emph{Quantum Gravity Partition Functions in Three Dimensions},
    \href{http://arxiv.org/abs/0712.0155}{\texttt{arXiv:0712.0155}}].
  \bibitem{mollifier} D. D. Ferrante, G. S. Guralnik, \emph{Mollifying Quantum
      Field Theory or Lattice QFT in Minkowski Spacetime and Symmetry Breaking},
    [\href{http://arxiv.org/abs/hep-lat/0602013}{\texttt{arXiv:hep-lat/0602013}}].
  \bibitem{symbtopind} D. D. Ferrante, G. S. Guralnik, \emph{From Symmetry
      Breaking to Topology Change I},
    [\href{http://arxiv.org/abs/hep-th/0609190}{\texttt{arXiv:hep-th/0609190}}].
  \bibitem{fredenhagen} Klaus Fredenhagen and Karl-Henning Rehren and Erhard
    Seiler, \emph{Quantum Field Theory: Where We Are},
    \href{http://www.arxiv.org/abs/hep-th/0603155}{\texttt{hep-th/0603155}}.
  \bibitem{reedsimon} M. Reed and B. Simon; \emph{Methods of modern
      mathematical physics}, Volume 1: Functional Analysis. Academic Press, New
    York 1972.
  \bibitem{dunfordschwartz} N. Dunford and J. T. Schwartz ; \emph{Linear operators},
    Volume 2: Spectral Theory. Interscience, New York 1964.
  \bibitem{colemandeluccia} Sidney Coleman, Frank De Luccia, \emph{Gravitational
      effects on and of vacuum decay},
    \href{http://prola.aps.org/abstract/PRD/v21/i12/p3305_1}{Phys. Rev. D 21,
      3305--3315 (1980)}.
  \bibitem{drg} Detlef Laugwitz, \emph{Differential and Riemannian Geometry}, Academic
    Press, 1965. Cornelius Lanczos,
    \emph{The Variational Principles of Mechanics}, Dover Publications, 1986.
    V. I. Arnold, \emph{Mathematical Methods of Classical Mechanics}. Springer,
    1997.
  \bibitem{seibergwitten} N. Seiberg, E. Witten, \emph{Monopoles, Duality and
      Chiral Symmetry Breaking in $N=2$ Supersymmetric QCD}, Nucl. Phys. B 431
    (1994) 484--550,
    \href{http://arxiv.org/abs/hep-th/9408099}{\texttt{hep-th/9408099}}.
  \bibitem{donaldson} S. K. Donaldson, \emph{The Seiberg-Witten equations and
      4-manifold topology},
    \href{http://www.ams.org/bull/1996-33-01/S0273-0979-96-00625-8/}{Bull. Amer. Math. Soc. 33 (1996), 45--70}.
  \bibitem{liviu} Liviu I. Nicolaescu, \emph{Notes on Seiberg-Witten
      Theory}. American Mathematical Society, 2000.
  \bibitem{geomana} J{\"u}rgen Jost, \emph{Riemannian Geometry and Geometric
      Analysis}. Springer.
  \bibitem{marek} Marek Szyd{\l}owski, Michale Heller, Wies{\l}aw Sasin,
    \emph{Geometry of spaces with the Jacobi metric},
    \href{http://scitation.aip.org/getabs/servlet/GetabsServlet?prog=normal&id=JMAPAQ000037000001000346000001&idtype=cvips&gifs=yes}{J. Math. Phys. 37
      (1), 1996, 346--360}.
  \bibitem{msheld} E. Witten, \emph{Mirror Symmetry, Hitchin's Equations, And
      Langlands Duality},
    \href{http://arxiv.org/abs/0802.0999}{\texttt{arXiv:0802.0999}}.
  \bibitem{lieairy} M. Kontsevich, \emph{Intersection theory on the moduli space
      of curves and the matrix Airy function},
    \href{http://projecteuclid.org/euclid.cmp/1104250524}{Comm. Math. Phys. Volume
      147, Number 1 (1992), 1-23}. R. N. Fernandez, V. S. Varadarajan,
    \emph{Airy Functions for Compact Lie Groups},
    [\href{http://arxiv.org/abs/0707.3235}{\texttt{arXiv:0707.3235}}].
\end{thebibliography}
\end{document}